\documentclass[journal]{IEEEtran}
\usepackage{hyperref}
\usepackage{geometry}
\usepackage{graphicx}
\usepackage{url}
\usepackage{cite}
\usepackage{longtable}
\usepackage{color,soul}
\usepackage{upgreek}
\setulcolor{red}
\sethlcolor{yellow}
\usepackage{cellspace, makecell, multirow}
\usepackage{subcaption}
\usepackage{amsmath,amsfonts,amssymb,bm}
\usepackage{amsthm}
\usepackage{multirow}
\usepackage{booktabs, multirow, array, makecell, caption}
\usepackage{algpseudocode}
\usepackage[vlined, ruled, boxed, linesnumbered]{algorithm2e}
\usepackage{indentfirst} 
\usepackage{breqn}

\usepackage{soul}
\let\existstemp\exists
\let\foralltemp\forall
\renewcommand*{\exists}{\existstemp\mkern2mu}
\renewcommand*{\forall}{\foralltemp\mkern2mu}
\newcommand\norm[1]{\left\lVert#1\right\rVert}
\DeclareMathOperator*{\argmin}{arg\,min}
\DeclareMathOperator*{\argmax}{arg\,max}
\theoremstyle{definition}

\begin{document}
\title{Optimization of Wireless Sensor Network Deployment for Spatiotemporal Reconstruction and Prediction}

\author{Jiahong~Chen,~\IEEEmembership{Member,~IEEE,}
       Teng~Li,~\IEEEmembership{Member,~IEEE,}
       Jing~Wang,
       and Clarence~W.~de~Silva,~\IEEEmembership{Fellow,~IEEE}
\thanks{J. Chen, T. Li, J. Wang, and C. W. de Silva are with the Industrial Automation Laboratory, Department of Mechanical Engineering, The University of British Columbia, Vancouver, BC, Canada. (e-mail: jhchen@mech.ubc.ca, tengli@mech.ubc.ca, j.wang94@alumni.ubc.ca, desilva@mech.ubc.ca. )}
\thanks{Manuscript received 20xx; revised 20xx.}}

\markboth{Journal of \LaTeX\ Class Files,~Vol.~xx, No.~x, Xxx~20xx}%
{Chen \MakeLowercase{\textit{et al.}}: Optimization of Wireless Sensor Network Deployment for Spatiotemporal Reconstruction and Prediction}
\maketitle

\begin{abstract}
This paper addresses the problem of optimizing sensor deployment locations to reconstruct and also predict a spatiotemporal field. A novel deep learning framework is developed to find a limited number of optimal sampling locations and based on that, improve the accuracy of spatiotemporal field reconstruction and prediction. The proposed approach first optimizes the sampling locations of a wireless sensor network to retrieve maximum information from a spatiotemporal field. A spatiotemporal reconstructor is then used to reconstruct and predict the spatiotemporal field, using collected in-situ measurements. A simulation is conducted using global climate datasets from the National Oceanic and Atmospheric Administration, to implement and validate the developed methodology. The results demonstrate a significant improvement made by the proposed algorithm. Specifically, compared to traditional approaches, the proposed method provides superior performance in terms of both reconstruction error and long-term prediction robustness.
\end{abstract}


\IEEEpeerreviewmaketitle

\section{Introduction}
\IEEEPARstart{A}{nalyzing} spatiotemporal fields (e.g., global sea surface temperature) plays an essential role in many areas of research, including climate change detection, weather forecasting, and water pollution mapping. Normally, a spatiotemporal field can be precisely sensed from satellites, using spaceborne imaging radars or thermal cameras. However, the accuracy and accessibility of the data obtained from these technologies may be affected by satellite limitations. For example, remotely sensed data may not be available in some areas due to weather conditions, such as cloud cover or a typhoon. Furthermore, only a limited types of sensing (e.g., temperature distribution, but not the pH value, conductivity, or oxidation-reduction potential) can be done by those methodologies. Therefore, it is essential to have alternative approaches to reconstruct and predict an entire spatiotemporal field when the satellite image is not accessible. Wireless sensor networks (WSNs), which may composed by boats and buoys, are robust to extreme weather and harsh environments; consequently, these networks can collect in-situ measurements in areas inaccessible to a satellite. Moreover, the spatiotemporal mappings between sparsely sampled measurements and an entire spatiotemporal field can be learned from historical satellite images. Hence, when a satellite image is unavailable, a spatiotemporal field can be reconstructed and predicted based on a limited amount of sampling data. Besides, these in-situ measurements can be regarded as a low-dimensional representation of the entire field. 

Usually, a spatiotemporal field can be encoded into a low-dimensional representation, which empowers sparse sampling. The low-dimensional sparse representation can then be used to reconstruct the original spatiotemporal field, which is a high-dimensional signal. Hence, there is a demand to optimize sensor deployment locations to estimate, predict, and control this high-dimensional signal. Notably, finding optimal measurement locations is intractable using brute-force approaches, which have been shown to be NP-hard. Therefore, the sampling locations of a WSN are usually chosen using convex optimization. For example, model-based approaches have been established for optimizing the measurement locations in moderate-sized spaces \cite{lan2016rapidly,  li2017automated, li2017hexagonal, chen2018rapidly}, and robotic buoys can be sent out for sampling and building the environment  \cite{nguyen2017adaptive, ma2017informative, ma2017data, dunbabin2017quantifying, hitz2017adaptive, chen2019deep}.However, model-based convex optimization methods highly rely on the environment model and only utilize a greedy approach to navigate robotic buoys to the most informative measurement locations. Hence, the model-based methods may fall in local minima and reduce the accuracy of spatiotemporal field reconstruction/prediction. Besides, the movement of buoys will consume considerable energy during the deployment.

Recently, data-driven approaches have been developed to optimize principal measurment locations. Data-driven approaches are of low complexity and easier to implement, compared to the model-based convex optimization approaches. Also, they can represent a massive region, e.g., $O(2^N)$, with merely $O(N)$ examples\cite{Goodfellow2016Deep}. Numerous complex signals can be reconstructed and predicted using data-driven sparse representations approaches, such as image reconstruction \cite{lustig2007sparse}, image denoising \cite{vincent2010stacked, burger2012image, xie2012image}, image super-resolution \cite{dong2014learning, dong2016image}, and structured signal recovery\cite{mousavi2015deep}.  Compressive sensing (CS) is a widely used signal recovery method for unknown signal reconstruction using undersampled sparse representations \cite{tipping2001sparse, zhang2014spatiotemporal}. Brunton et al., propose a compressive sensing-based sparse sampling method for characterizing or classifying a high-dimensional system \cite{brunton2016sparse}. The proposed algorithm solves the $\ell_1$ minimization of finding the measurement with fewest zero entities for reconstructing the original signal. Lu et al. propose a convolutional CS framework to avoid the inefficiency and blocking artifacts in the traditional CS algorithms \cite{lu2018convcsnet}. Their algorithm senses the input image using a set of convolutional filters, and reconstructs the image using linear convolutional measurements.

Although CS can recover a wider class of signals, it has limitations in capturing spatiotemporal patterns. CS-based algorithms have difficulties in obtaining nonlinear spatiotemporal mappings, and do not perform well when the volume of the signal is large. Besides, CS-based algorithms need to solve a linear programming problem to recover the low-dimensional signal, which has low efficiency and is time consuming.

In constrast to CS-based approaches, principal component analysis (PCA) computes low-dimensional patterns and features directly from high-dimensional signals. PCA transforms the input signal into an orthogonal coordinate system and projects the signal onto the coordinates according to the significance of the variance. Then, the selected principal conponents retain the most variance in the input signal, and can be used to reconstruct the system rather easily. Manohar et al. propose an optimal sparse samplnig scheme based on the matrix QR factorization and singular value decomposition (SVD) \cite{manohar2018data}, which outperforms the CS-based approaches. Guo et al. develop a sparse sampling based topological data analysis technique to reconstruct a high-dimensional signal from limited observations \cite{guo2018sparse}. The optimal sparse measurements are selected and used for reconstructing the input signal at high efficiency and precision. Lu et al. use SVD to learn basis from training data, and reconstruct fluid flows using random sample \cite{lu2018interplay}. They investigate the interplay of data sparsity in the underlying flow system and reconstruct the entire spatiotemporal field using limited in situ measurements.

Deep learning (DL) methods can also be used for signal compression and reconstruction. In contrast to CS-based and PCA-based approaches that employ only linear sparse representations for signal reconstruction, DL methods can support both linear and nonlinear reconstruction. Normally, deep generative models encode an input signal into a latent low-dimensional space, which can reconstruct the high-dimensional siganl. They are widesly utilized in many fields such as image recognition  \cite{krizhevsky2012imagenet}, recommendation systems \cite{Tang2018personalized}, power management \cite{shu2017energy}, model compression \cite{tang2018ranking} and data encoding/decoding \cite{kingma2013auto, goodfellow2014generative}. Kingma et al. develop variational autoencoders (VAEs) to learn the low-dimensional latent representation of an input signal  \cite{kingma2013auto}. Similarly, generative adversarial network (GAN) can produce superficially real images from low-dimensional latent space  \cite{goodfellow2014generative}. Furthermore, DL methods are also well-suited for making predictions based on temporally correlated signal. Recurrent neural networks, such as long short-term memory (LSTM), exploit temporal dynamics of the input signal to give roubust prediction on time series data \cite{qin2017dual, shu2018energy}. However, traditional DL methods are not suitable for the indicated task because the latent low-dimensional space cannot be sampled as in-situ measurements.

\begin{figure}
\centering
\includegraphics[width=0.45\textwidth]{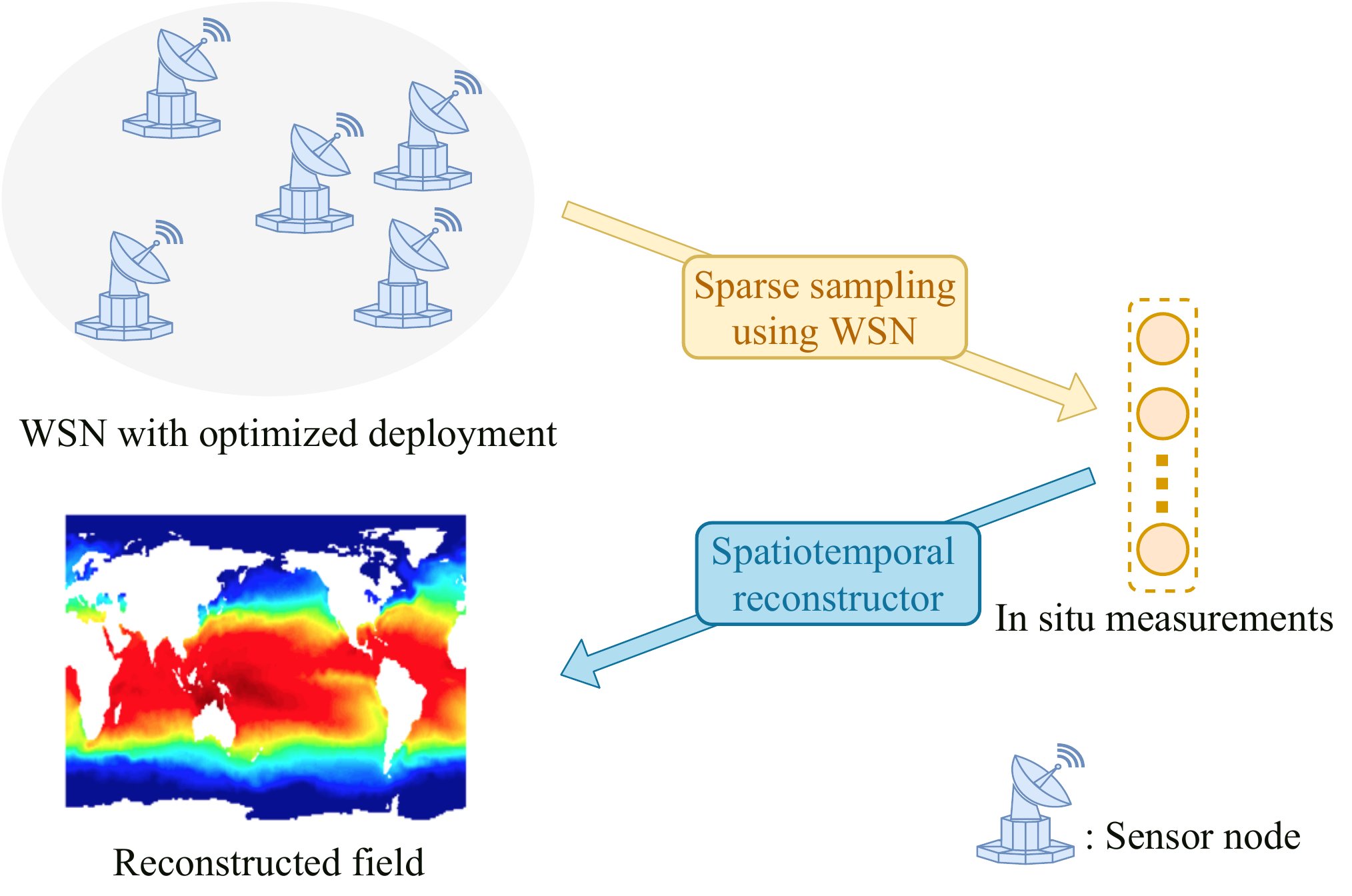} 
\caption{Architecture of the proposed WSN-based reconstruction system}\label{fig:architecture}
\end{figure}

In the present paper, a novel DL model, is proposed to optimize the sensor deployment locations and reconstruct a spatiotemporal field from limited in-situ measurements. The spatiotemporal field is encoded into sparse sampling locations where the WSN is deployed, and the in-situ measurements collected by the WSN are regarded as a low-rank representation of the entire spatiotemporal field. A spatiotemporal reconstructor then captures the nonlinear spatiotemporal mappings between the sparse sampling locations and the rest of the field. Therefore, the in-situ measurements collected at the optimized sampling locations can be reconstructed as the entire spatiotemporal field. The systemic design of the WSN-based reconstruction system is shown in Figure \ref{fig:architecture}.

The contributions of the present paper are summarized as follows:
\begin{itemize}
\item The deployment of a WSN is optimized for spatiotemporal field reconstruction and prediction.
\item A novel DL framework is developed to reconstruct a spatiotemporal field from limited in-situ measurements sampled by a WSN.
\item Reconstruction accuracy and long-term prediction roubustness are significantly improved in both sea surface temperature (SST) dataset and global precipitation dataset.
\end{itemize}

The rest of the paper is organized as follows. Section \ref{section:preliminary} discusses the preliminaries and formulates the problem. Section \ref{section:sensor_selection} presents the scheme for sampling location optimization and the DL model for reconstructing spatiotemporal field. The performance and the efficiency of the proposed algorithm are evaluated in Section \ref{section:simulation_result_generative} via extensive simulation results. Conclusions of the paper are given in Section \ref{section:futurework_generative}.

\section{Preliminaries and Problem Formulation}\label{section:preliminary}
There are $\binom{m}{r}=\frac{m(m-1)\dots (m-r+1)}{r(r-1)\dots 1}$ possible selections of $r$ sensors in an $m$-dimensional environment. 
It has been shown that recovering the $m$-dimensional signal from $r\ll m$ observations is an NP-hard problem \cite{baron2009distributed}. 
Nevertheless, the PCA-based methods have shown their capability to recover high-dimensional signals from a limited information content.

PCA can represent a high-dimensional signal $\boldsymbol{\phi} \in \mathbb{R}^m$ as a linear combination of orthogonal eigenmodes (modal vectors). Hence, the high-dimensional signal can be projected onto the lower-dimensional PCA subspace, using singular value decomposition (SVD).

Given measurements $\boldsymbol{\Phi} = [\boldsymbol{\phi}_1 \boldsymbol{\phi}_2 \dots \boldsymbol{\phi}_M]$ for $M$ snapshots of the input signal, it can be represented by the orthonormal left singular vectors $\boldsymbol{\Psi}$, right singular vectors $\mathbf{V}$ and the diagonal matrix $\boldsymbol{\Sigma}$:
\begin{equation} \label{SVD-equation}
\boldsymbol{\Phi} = \boldsymbol{\Psi} \cdot \boldsymbol{\Sigma} \cdot \mathbf{V}.
\end{equation}

Then, the dimension of the right hand part of Equation \ref{SVD-equation} can be reduced to $r$ according to Eckart-Young theorem \cite{eckart1936approximation}:
\begin{equation} \label{SVD-result-of-original-signal}
\begin{split}
&\boldsymbol{\Phi} \approx \boldsymbol{\Phi}^* = \argmin_{\tilde{\boldsymbol{\Phi}}} \norm{\boldsymbol{\Phi} - \tilde{\boldsymbol{\Phi}}}_F\\
s.t. ~~~&rank(\tilde{\boldsymbol{\Phi}})=r,
\end{split}
\end{equation}
where $\boldsymbol{\Phi}^* =  \boldsymbol{\Psi}_r \cdot \boldsymbol{\Sigma}_r \cdot \mathbf{V}_r$ and $\norm{\cdot}_F$ is the Frobenius norm. Besides, matrices $\boldsymbol{\Psi}_r$ and $\boldsymbol{\Sigma}_r$ are the first $r$ rows and columns of $\boldsymbol{\Psi}$ and $\boldsymbol{\Sigma}$, respectively. And $\boldsymbol{\Sigma}_r$ is the first $r\times r$ block of $\boldsymbol{\Sigma}$. Rank $r$ is usually chosen by filtering singular values to capture most variances in the dataset while not magnifying the noise \cite{gavish2014optimal}. Hence, PCA can reduce the dimension of the high-dimensional signal by using orthogonal projection.

However, PCA-based approaches have limitations in extracting the nonlinear mappings between the low-dimensional representations and the high-dimensional space. It only applies linear matrix multiplication to reconstruct the spatiotemporal field. In constrast, DL approaches can learn the nonlinear spatiotemporal pattern of a signal by extracting dominating features from the training set. Hence, a higher reconstruction and prediction performance can be achieved.

In the present paper, a DL model is proposed to reconstruct a full signal from a small amount of measurements. Denote the physical phenomenon of a spatiotemporal field as the high-dimensional signal $\boldsymbol{\phi}\in \mathbb{R}^m$. The nonlinear spatiotemporal dynamics of $\boldsymbol{\phi}$ can be captured by the low-dimensional measurements. Hence, $\tilde{\boldsymbol{\phi}}$ can be reconstructed from its low-rank representations based on sparsely sampled in situ measurements:
\begin{equation}\label{low-rank-representation}
\tilde{\boldsymbol{\phi}} = \mathcal{G}_r(\bm{\uptheta}|\mathbf{y}),
\end{equation}
where $\mathcal{G}_r:\mathbb{R}^r \rightarrow \mathbb{R}^m$ is the nonlinear reconstruction function, $\bm{\uptheta}$ is the model parameter, and $\mathbf{y}\in \mathbb{R}^r$ is the sparsely sampled measurement from $\boldsymbol{\phi}$. 

In the present paper, the sparsely sampled in situ measurements, $\mathbf{y}$, are encoded from the entire field, $\boldsymbol{\phi}$, using a measurement matrix $\mathbf{C}$. The measurement matrix can be used to decide deployment locations and extract in situ measurements directly from the spatiotemporal field:
\begin{equation}\label{optimal-measure-for-reconstruction}
\mathbf{y} = \mathbf{C}\cdot \boldsymbol{\phi}.
\end{equation}

Therefore, the principal task of the proposed work is two-fold. First, the measurement matrix $\mathbf{C} \in \mathbb{R}^{r\times m}$ should be optimized to sparsely sampled critical observations from $\boldsymbol{\phi}$. Second, a suitable spatiotemporal reconstructor should be designed to learn the nonlinear reconstruction function $\mathcal{G}_r$.


\section{WSN Deployment Optimization for Spatiotemporal Reconstruction}\label{section:sensor_selection}
In this section, a novel DL model is proposed to optimize sampling locations that can best represent the spatiotemporal field. It utilizes statistical approches to find key sampling locations and incorporates a spatiotemporal reconstructor to learn the nonlinear mappings between in situ measurements and the entire spatiotemporal field. In this manner, the WSN can be deployed to the selected sampling locations, and the spatiotemporal field can be reconstructed and predicted by using only the sampled in situ measurements.

\subsection{Sparse Sampling for Reconstruction}
In the present work, a high-dimensional signal is reconstructed using limited in situ measurements. With that objective, the sampling locations are optimized to achieve better reconstruction performance. Denote the spatiotemporal field as an $m$-dimensional space:
\begin{equation}
  \boldsymbol{\phi} = [\phi_{1} \phi_{2} \cdots \phi_{m}]^T,
\end{equation}
where $\phi_{\cdot}$ is a random variable that represents the sampling value at the corresponding location. Then, the measurement matrix $\mathbf{C}$ can be used to select the sampling locations:
\begin{equation}
\mathbf{C}=[\mathbf{e}_{\gamma_1} \mathbf{e}_{\gamma_2} \cdots \mathbf{e}_{\gamma_r}]^T,
\end{equation}
where $\mathbf{e}_{\cdot} \in \mathbb{R}^m$ are the one-hot canonical basis vectors.

According to Equation \ref{optimal-measure-for-reconstruction}, observations collected by the WSN can be expressed as a linear combination of the canonical basis and the high dimensional signal:
\begin{equation}
y_i = \mathbf{e}_{\gamma_i}\boldsymbol{\phi}
\end{equation}

Then, the observation $\mathbf{y}$ can be simplified as:
\begin{equation}
\mathbf{y} = [\phi_{\gamma_1} \phi_{\gamma_2} \cdots \phi_{\gamma_r}]^T,
\end{equation}
where $\gamma = \{\gamma_1, \gamma_2, \cdots, \gamma_r \} \subset \{1, 2, \cdots, m\}$ is the set of indices for the selected sampling locations.  Hence, $\mathbf{y}$ is a subset of the $m$-dimensional spatiotemporal field, and can be regarded as the sparesly sampled in situ measurements. Then, as $\mathbf{y}$ can be directly observed from the spatiotemporal field, the spatiotemporal field $\tilde{\boldsymbol{\phi}}$ can be reconstructed, given the observations at the sampling locations and the nonlinear mapping $\mathcal{G}_r$.

\begin{figure}
\centering
\includegraphics[width=0.4\textwidth]{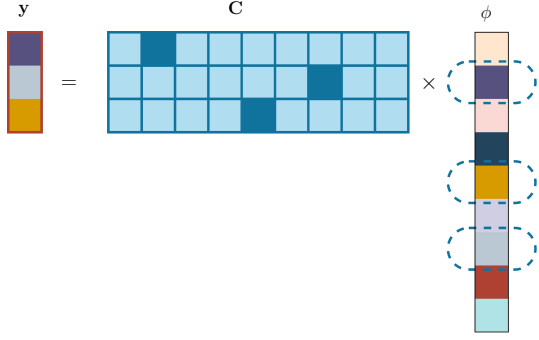} 
\caption{Illustration of generating sparse sampling locations using a canonical basis. Dark blue region in matrix $\mathbf{C}$ stands for 1, rest of the matrix contains 0; second, seventh and fifth row of $\boldsymbol{\phi}$ are extracted according to $\mathbf{C}$, to compose $\mathbf{y}$.}\label{fig:C-matrix}
\end{figure}

A schematic diagram for sparse sampling is given in Figure \ref{fig:C-matrix}. The observations at the sampling locations $\mathbf{y}$ are selected by the measurement matrix $\mathbf{C}$ to guarantee the best feasible reconstruction $\tilde{\boldsymbol{\phi}}$, and the sampling locations in the spatiotemporal field correspond to the in situ measurements in the input matrix $\boldsymbol{\phi}$. Hence, the sparse sensor deployment strategy should compute the rows of $\boldsymbol{\phi}$ that optimally represent the spatiotemporal mappings in the field. Then, $\mathcal{G}_r$ should be trained to learn spatiotemporal mappings between the observation $\mathbf{y}$ and $\boldsymbol{\phi}$. 
 
In the present paper, a data-driven approach is utilized to find optimal sampling locations and the canonical matrix $\mathbf{C}$.
According to Equation \ref{SVD-result-of-original-signal}, the principal component transformation of $\boldsymbol{\Phi}$ can be accomplished by matrix factorization techniques, such as SVD. The score matrix $\mathbf{T}$ can be represented as: 
\begin{equation}
    \begin{split}
        \mathbf{T}&= \boldsymbol{\Phi}\mathbf{V}^T\\
        &= \boldsymbol{\Psi} \boldsymbol{\Sigma} \mathbf{V}\mathbf{V}^T\\
        &= \boldsymbol{\Psi} \boldsymbol{\Sigma}.\\
    \end{split}
\end{equation} 

Only $r$ principal components needs to be kept:
\begin{equation}\label{principal-components}
\begin{split}
    \mathbf{T}_r&= \boldsymbol{\Psi}_r \cdot \boldsymbol{\Sigma}_r\\
    &=\boldsymbol{\Phi}\mathbf{V}^T_r.\\
\end{split}
\end{equation} 

Therefore, denoting $\mathbf{T}_r = \boldsymbol{\Psi}_r \cdot \boldsymbol{\Sigma}_r$ as the principal basis, the input data $\boldsymbol{\Phi}$ can be reconstructed \cite{shlens2014tutorial}. Then, optimal sampling locations can be computed with regard to the principal basis $\mathbf{T}_r$. According to Equation \ref{low-rank-representation}, input signal $\boldsymbol{\phi}$ can be compressed as $\mathbf{y} = \mathbf{C}\boldsymbol{\phi}$. Hence, the input training dataset can be expressed as $\boldsymbol{\Phi}$ and the corresponding sparesely sampled signal is $\mathbf{Y} = [\mathbf{y}_1 \mathbf{y}_2 \dots \mathbf{y}_M]$. Given a suitable canonical matrix $\mathbf{C} \in \mathbb{R}^{r\times m}$, the input data matrix can be compressed to a limited number of in situ measurements: 
\begin{equation}\label{compression-from-original-signal}
\mathbf{Y} = \mathbf{C}\boldsymbol{\Phi}
\end{equation}

Substitute Equation \ref{principal-components} into Equation \ref{compression-from-original-signal}:
\begin{equation}
\begin{split}
    \mathbf{T}_r & = \boldsymbol{\Phi}\mathbf{V}^T_r\\
    \mathbf{T}_r & = (\mathbf{C}^{-1}\mathbf{Y})\cdot \mathbf{V}^T_r\\
    (\mathbf{V}^T_r)^{-1} \cdot \mathbf{T}_r & = \mathbf{C}^{-1}\mathbf{Y}\\
    \mathbf{V}_r& = \mathbf{T}_r^{-1} \mathbf{C}^{-1}\mathbf{Y}\\
\mathbf{V}_r&=(\mathbf{C}\mathbf{T}_r)^{-1}\mathbf{Y}.
\end{split}
\end{equation}

Therefore, the spatiotemporal field $\tilde{\boldsymbol{\Phi}}$ can be represented using a limited number of measurements $\mathbf{Y}$:
\begin{equation}\label{optimize-canonical-matrix}
    \begin{split}
        \tilde{\boldsymbol{\Phi}} & = \boldsymbol{\Psi}_r \cdot \boldsymbol{\Sigma}_r \cdot \mathbf{V}_r\\
        & = \mathbf{T}_r \cdot \mathbf{V}_r\\
        &= \mathbf{T}_r(\mathbf{C}\mathbf{T}_r)^{-1}\mathbf{Y}.
    \end{split}
\end{equation}

As shown in Equation \ref{optimize-canonical-matrix}, the reconstruction of  $\tilde{\boldsymbol{\Phi}}$ depends on the canonical basis $\mathbf{C}$. Hence, the optimization of sampling locations ($\hat{\gamma}$) for reconstruction also requires the optimization of the measurement matrix $\mathbf{C}_\gamma$. Therefore, the sampling locations of the WSN are optimized by maximizing the singular value spectrum of the principal components:
\begin{equation}\label{singular-value-spectrum-maximization}
\hat{\gamma} = \argmax_\gamma{|\det \mathbf{C}_\gamma\mathbf{T}_r|}.
\end{equation}

The optimization of Equation \ref{singular-value-spectrum-maximization} can be done by heuristic greedy approaches, such as Karhunen-Lo\`{e}ve transform \cite{everson1995karhunen} and empirical interpolation methods (EIMs) \cite{drmac2016new, chaturantabut2010nonlinear}. In the present paper, the optimization of Equation \ref{singular-value-spectrum-maximization} is carried out by calculating the QR decomposition of $\mathbf{T}_r$. The optimal sampling locations, $\gamma$, correspond to the top-$r$ pivots in the result of the QR decomposition.

Algorithm \ref{algo:sparse-sampling} presents the calculation of optimal sampling locations. It first computes the principal components, in line 1 and line 2, from the input data. Then, $r$ sampling locations are selected iteratively, from line 4 to line 9. In each iteration, the row with the maximum norm is selected as the pivot for sampling, using Householder relections 
\cite{manohar2018data}. Then, the measurement matrix C can be calculated based on the output of the Algorithm 1.

\begin{algorithm}
\SetAlgoLined
\KwIn{$\boldsymbol{\Phi},r$}
\KwOut{$\gamma$ }
$\boldsymbol{\Psi}_r, \boldsymbol{\Sigma}_r, \mathbf{V}_r$ = SVD($\boldsymbol{\Phi}$)\;
$\mathbf{T}_r = \boldsymbol{\Psi}_r\cdot \boldsymbol{\Sigma}_r$\;
$\gamma\leftarrow \emptyset$\;
\For{i = 1 \textbf{to} r}{
	$\tilde{\gamma} = \argmax_{\tilde{\gamma}\notin\gamma}{\norm{\mathbf{t}_{\tilde{\gamma}}}}, \mathbf{t}_i := \textrm{row}_i\mathbf{T}_r$\;
	$\mathbf{T}_r \leftarrow swap(\mathbf{t}_i,\mathbf{t}_{\tilde{\gamma}})$\;
	Apply Householder reflections\;
	$\gamma = \gamma\cup\tilde{\gamma}$\;
}
\caption{Sparse Sampling for Reconstruction}
\label{algo:sparse-sampling}
\end{algorithm}

Then, the computed deployment locations $\gamma$ can establish a WSN that collects in situ measurements from the spatiotemporal field. Also, it is essential to preserve the network connectivity during the process of in situ measurement sampling. The complete network connectivity ensures that the sensor nodes can communicate with each other, and the collected data can be transmitted to the data center. 

In the present paper, several assumptions are made for analyzing the network connectivity of WSN. First, the calculation of the network connectivity is only based on the location of the sensor nodes and does not depend on the size of the sensor node. Second, the bandwidth of the WSN is sufficient for data transmission and sensor node communication. Third, all the sensors are assumed to be homogeneous. Then, a WSN is said to be connected if every sensor node has at least one adequately close neighbor:
\begin{equation}
\forall \gamma_i \in \gamma, \exists \gamma_j \in \gamma: \norm{\gamma_i - \gamma_j}\leq \tau_{com} (i \neq j).
\end{equation}
where $\tau_{com}$ is the threshold of communication distance, and $\norm{\cdot}$ is the $\ell_1$ norm. 

In other words, the WSN is said to be connected if the distance between the sensor node $\gamma_i$ and its closest distance is within $\norm{\cdot}$. Denote the closest neighbor distance of $\gamma_i$ as $\omega_i = \min_{\gamma_k \in \gamma}{\norm{\gamma_k-\gamma_i}}$, which also represents the minimal communication distance required by $\gamma_i$.  Hence, the communication threshold of WSN should be at least $\max_{i=1,2,\cdots, r}{\omega_i}$ to ensrue that all sensor nodes are connected. The minimal communication distance required by the WSN can be computed as:
\begin{equation}
\Omega = \max_{i=1,2,\cdots, r}{\omega_i} = \max_{\gamma_i\in \gamma}(\min_{\gamma_k \in \gamma}{\norm{\gamma_k-\gamma_i}}).
\end{equation}

Thus, if $\Omega\leq \tau_{com}$, the WSN is connected. Otherwise, bridging sensor nodes should be added accordingly, to ensure the network connectivity of $\gamma$.

\subsection{Deep Learning based Spatiotemporal Field Reconstruction}
Although Euqation \ref{optimize-canonical-matrix} can represent the spatiotemporal field $\boldsymbol{\Phi}$ from sampling data $\mathbf{Y}$, it only captures the linear correlation and may result in low accuracy for complex signals. Therefore, a DL-based spatiotemporal reconstructor is proposed in this section to learn the nonlinear spatiotemporal projection from the sampling data and the spatiotemporal field.

A deep multilayer preceptron is proposed to learn the signal reconstruction from limited in situ measurements. In the beginning, the input signal is compressed to $r$-dimensional measurements using matrix $\mathbf{C}$:
\begin{equation}
\mathbf{y} = \mathbf{C}\cdot \boldsymbol{\phi}.
\end{equation}

\begin{figure}
\centering
\includegraphics[width=0.4\textwidth]{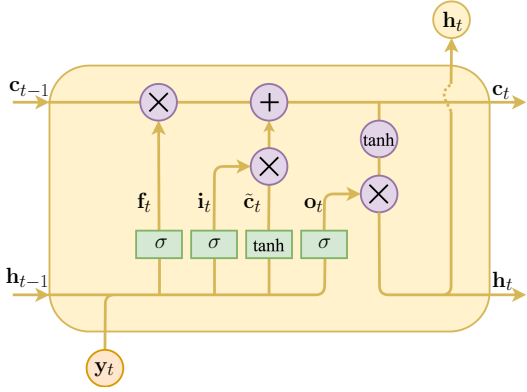} 
\caption{Details of the LSTM cell. Purple circles indicates the pointwise operation; green rectangles are the neural network layers.}\label{fig:LSTM-cell}
\end{figure}

Next, the input signal is passed through the LSTM cell to learn the inherent temporal correlation of the input time series $\mathbf{y}_t$. Details of the LSTM cell is shown in Figure \ref{fig:LSTM-cell}. The key idea about the LSTM is the repeating module that continously update cell state vector $\mathbf{c}_t$ and hidden state vector $\mathbf{h}_t$. In the beginning, forget gate layer $\mathbf{f}_t$ forgets part of information in input $\mathbf{y}_t$ and hidden state $\mathbf{h}_t$, using sigmoid function \cite{hochreiter1997long}:
\begin{equation}
\mathbf{f}_t=\sigma(\mathbf{W}_f [\mathbf{h}_{t-1}, \mathbf{y}_t ]+\mathbf{b}_f ),
\end{equation}
where $\sigma(\cdot)$ is the sigmoid function, $\mathbf{W}_f$ is the weight matrix and $\mathbf{b}_f$ is the bais. Next, new information for updating cell state in time $t$ is captured, using both sigmoid function and hyperbolic tangent function:
\begin{equation}
\begin{split}
\mathbf{i}_t&=\sigma(\mathbf{W}_i [\mathbf{h}_{t-1}, \mathbf{y}_t ]+\mathbf{b}_i )\\
\tilde{\mathbf{c}}_t&=\tanh(\mathbf{W}_c [\mathbf{h}_{t-1}, \mathbf{y}_t ]+\mathbf{b}_c),
\end{split}
\end{equation}
where $\tanh(\cdot)$ is the hyperbolic tangent function $\mathbf{W}_\cdot$ is the weight matrix and $\mathbf{b}_\cdot$ is the bais. After this, the cell state is updated using $\mathbf{f}_t$, $\mathbf{i}_t$ and $\tilde{\mathbf{c}}_t$:
\begin{equation}
\mathbf{c}_t = \mathbf{f}_t\cdot \mathbf{c}_{t-1} + \mathbf{i}_t \cdot \tilde{\mathbf{c}}_t 
\end{equation}

The last step of the LSTM cell updates the hidden state, which is based on output gate vector $\mathbf{o}_t$ and updated cell state $\mathbf{c}_t$:
\begin{equation}
\begin{split}
\mathbf{o}_t &= \sigma(\mathbf{W}_o[\mathbf{h}_{t-1},\mathbf{y}_t]+\mathbf{b}_o)\\
\mathbf{h}_t &= \mathbf{o}_t\times \tanh(\mathbf{c}_t),
\end{split}
\end{equation}
where $\mathbf{W}_o$ is the weight matrix and $\mathbf{b}_o$ is the bais.

\begin{figure*}
\centering
\includegraphics[width=0.8\textwidth]{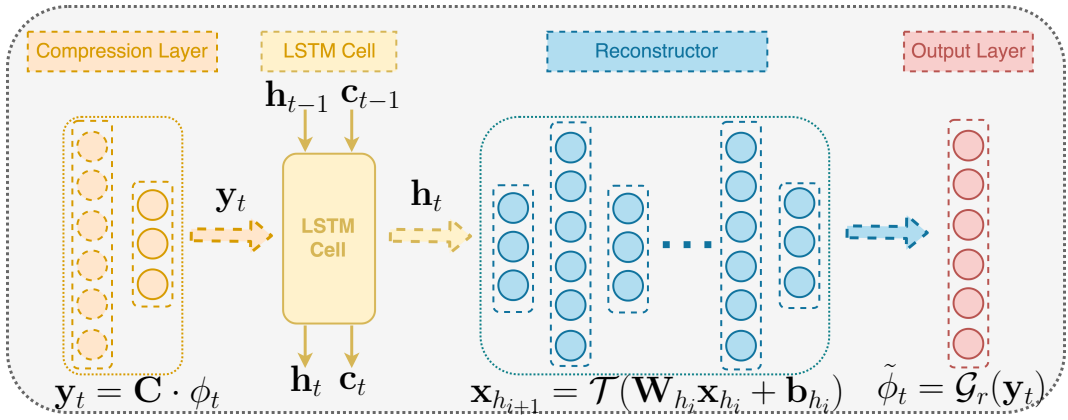} 
\caption{Transitions between different modules of the proposed model. Orange module represent the compression layers that output optimized measurement; yellow module is the LSTM cell that stores temporal information; blue module are neurons in the hidden layer for spatiotemporal field reconstruction; red module is the output layer.}\label{fig:cell-transitions}
\end{figure*}

Then, the output of LSTM cell can be passed to the reconstructor to learn the spatiotemporal mappings between the in situ measurements and a entire spatiotemporal field. The transitions between the modules of the proposed model is presented in Figure \ref{fig:cell-transitions}. Two-dimensional images of the spatiotemporal field are reshaped into a one-dimensional vector at size $m$ and fed into the compression layer. Then, $r$ key sampling locations, $\mathbf{y}$, are selected via the canonical measurement matrix $\mathbf{C}$. Next, $\mathbf{y}$ is passed through the multi-layer reconstructor to reconstruct $\tilde{\boldsymbol{\phi}}$. All hidden layers are applied with nonlinear Rectified Linear Units (ReLU). Last, Adam optimization is utilized to compare the difference between $\tilde{\boldsymbol{\phi}}$ and $\boldsymbol{\phi}$, and minimize the loss.

Hence, each layer of the reconstructor can be represented as:
\begin{equation}
  \mathbf{x}_{h_{i+1}} = \mathcal{T}(\mathbf{W}_{h_{i}}\mathbf{x}_{h_{i}}+\mathbf{b}_{h_{i}}),
\end{equation}
where $\mathbf{W}_{h_{i}} \in \mathbb{R}^{m\times r}, \mathbf{b}_{h_{i}}\in \mathbb{R}^m$ or $\mathbf{W}_{h_{i}} \in \mathbb{R}^{r\times m}, \mathbf{b}_{h_{i}}\in \mathbb{R}^r$, depending on the size of the input data. Besides, $\mathbf{x}$ denotes the hidden neurons and $\mathcal{T}(\cdot)$ is the element-wise nonlinear rectifier function. Note that the input of the reconstructor, $\mathbf{x}_{h_{0}}$, is $\mathbf{h}_t$. Then, the calculation for the output layer is given by:
\begin{equation}
\tilde{\boldsymbol{\phi}} = \mathcal{T}(\mathbf{W}_r\mathbf{x}_h+\mathbf{b}_r),
\end{equation}
where $\mathbf{W}_r \in \mathbb{R}^{r\times m}$ and $\mathbf{b}_r\in \mathbb{R}^m$ are the weighting matrix and the bias of the last hidden layer, respectively. As the size of the output layer must match the dimension of the environment, the number of hidden layers must be even.

Denote the trainable variables as $\bm{\uptheta} = \{\mathbf{W}_{h_1}$, $\mathbf{b}_{h_1}$, $\mathbf{W}_{h_2}$, $\mathbf{b}_{h_2}$, $\cdots$, $\mathbf{W}_{r}$, $\mathbf{b}_{r}$, $\mathbf{W}_{f}$, $\mathbf{b}_{f}$, $\mathbf{W}_{i}$, $\mathbf{b}_{i}$, $\mathbf{W}_{o}$, $\mathbf{b}_{o}$, $\mathbf{c}$, $\mathbf{h}\}$.  The nonlinear mapping from the sprase observations to the output data $\tilde{\boldsymbol{\phi}}$ can be represented as $\tilde{\boldsymbol{\phi}} = \mathcal{G}_r(\bm{\uptheta}|\mathbf{y})$. The mean squared error (MSE) is used as the loss function: 
\begin{equation}
\mathcal{L}(\bm{\uptheta}|\mathbf{y}) = \frac{1}{M}\sum_{i=1}^M\norm{\mathcal{G}_r(\bm{\uptheta}| \mathbf{y}_i)-\boldsymbol{\phi}_i}^2_2
\end{equation}

\begin{figure}
\centering
\includegraphics[width=0.5\textwidth]{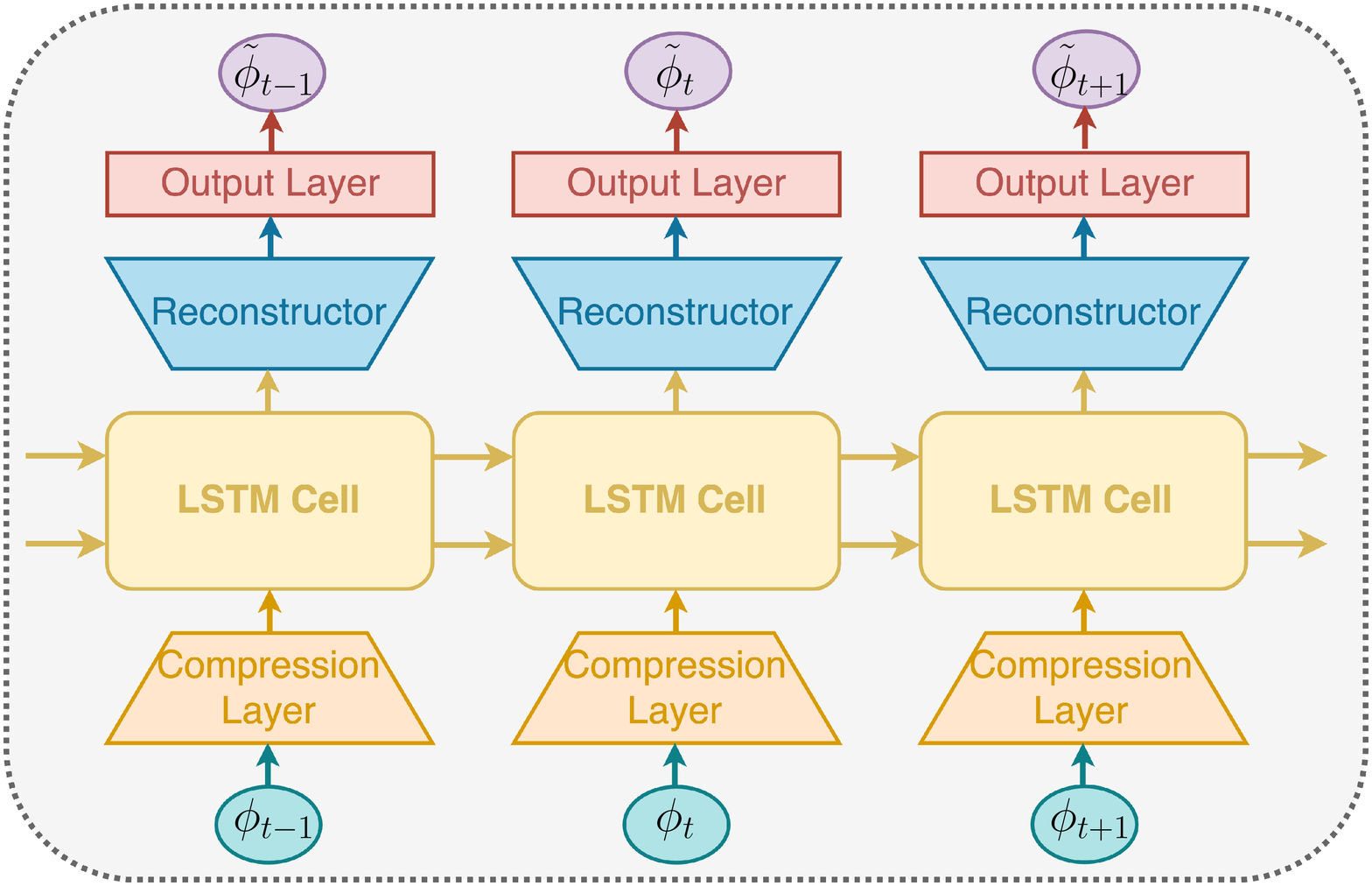} 
\caption{The architecture of the proposed model.}\label{fig:model-framework}
\end{figure}

The entire framework of the proposed model is shown in Figure \ref{fig:model-framework}. At each time step, input signals are fed into the compression layer to obtain the in situ measurements. Then, the collected measurements are passed to the LSTM cell, which learns the temporal correlation of the time series data and connects the reconstructor. Note that the information learned at time $t$ will be passed to the LSTM cell at time $t+1$. These repeated LSTM cells establishes a chain structure and updates itself across the entire temporal space, which can capture both long term and short term temporal information. After this, the reconstructor utilizes the output of the LSTM cell to reconstruct the spatiotemporal field $\tilde{\boldsymbol{\phi}}$.

Once the training process minimizes the difference between the model output and the input data, the spatiotemporal reconstructor learns the nonlinear mapping between the measurements of the selected deployment locations and the entire spatiotemporal field. Hence, once new data are sampled by the WSN, the entire spatiotemporal field can be reconstructed.

The training set for the proposed DL model has $M$ pairs of the input signal and the corresponding observations in the selected locations: 
\begin{displaymath}
  \mathcal{S}_{train} = \{(\mathbf{y}_1, \boldsymbol{\phi}_1), (\mathbf{y}_2, \boldsymbol{\phi}_2), \cdots, (\mathbf{y}_M, \boldsymbol{\phi}_M)\}.
\end{displaymath} 

Similarly, the test set $\mathcal{S}_{test}$ contains several pairs of signals and observations.

Algorithm \ref{SSS-main} describes how the spatiotemporal projection between the sampling data and the entire field is learned.

\begin{algorithm}
    \SetAlgoLined
    \KwIn{$\mathcal{S}_{train}$, $\boldsymbol{\gamma}$, $epoch$, $batch\_size$}
    \KwOut{$\bm{\uptheta}$ }
    Initialization: randomly draw $\bm{\uptheta}$ from uniform distributions with a zero mean.\;
    $\boldsymbol{\Phi}_b = [~]$, $\mathbf{C} = zeros(r,m)$\;
    \For{$i = 1 $ \textbf{to} $ r$}{
        $\mathbf{C}[i,\gamma_r] = 1$\;
    }
    $\mathcal{S}_{train} \leftarrow$ normalize $\mathcal{S}_{train}$\; 
    \For{ $i$ =1 \textbf{to} $epoch$}{
        $idx = (i-1)\times batch\_size$\;
        \For {$j$ = 1 \textbf{to} $batch\_size$}{
            $\boldsymbol{\Phi}_b\ \leftarrow 
            \begin{bmatrix}
                \boldsymbol{\Phi}_b & \boldsymbol{\phi}_{idx+j}
            \end{bmatrix}$ \;
        }
         
        Use Adam optimization to minimize $\mathcal{L}(\mathbf{y}| \bm{\uptheta})$ with input $\boldsymbol{\Phi}_b$\
    }
    \caption{The proposed DL model}
    \label{SSS-main}
\end{algorithm}


Initially, the hyper-parameters and trianable variables are initialized. Then, line 3 to line 5 generate the measurement matrix $\mathbf{C}$ from the optimized sampling locations $\gamma$. Next, the training process iterates for $epoch$ times to learn the reconsturction model $\mathcal{G}$. In each iteration, a batch of training data $\boldsymbol{\Phi}_b$ is selected from $\mathcal{S}_{train}$ between line 7 and line 11. Last, line 12 minimize the model loss according to the model indicated in Figure \ref{fig:model-framework}.

\section{Simulation and Results}\label{section:simulation_result_generative}
This section presents  numerical simulations using data from the National Oceanic and Atmospheric Administration (NOAA). The algorithm developed in the present work is compared with some state-of-the-art algorithms for reconstructing and predicting complex spatiotemporal fields.

\subsection{Experimental Setup}\label{subsection:GEN-Experiment-setup}

\subsubsection{Model training} 
Back-propagation is used to fine-tune the layer weights and biases for the DL models. The proposed network is trained with Adam optimizer  \cite{kingma2014adam} by using the default settings: $\beta_1 = 0.9$, $\beta_2 = 0.999$ and $\epsilon = 10^{−8}$. The learning rate is set to 0.001 with a cosine decay, and the mini-batch size is set to 20. The training metrics are partially adapted from \cite{he2019bag}. The dataset is divided into a training set and a test set, according to their sequence in the time series. The beginning part of the time series is regarded as the training set for learning the nonlinear map from the observations to the entire spatiotemporal field. The test set then comprises the remaining future snapshots. In this manner, given the future in situ measurements, the ability to predict the future spatiotemporal field can be examined as well. The simulation was conducted on the Google cloud platform with Dual-Core CPU, 13 GB memory and a Nvidia Tesla K80 GPU. The DL model was implemented using the TensorFlow framework \cite{tensorflow2015-whitepaper}.

\subsubsection{Dataset}
Simulations were conducted using two global climate datasets. The first one is the NOAA global SST dataset spanning from 1990 to 2018, which is publicly available online at \cite{NOAA-sst}. The SST dataset provides weekly global sea surface temperature means in 1.0-degree latitude $\times$ 1.0-degree longitude global grid (180 $\times$ 360) \cite{reynolds2002improved}. The second dataset is the NOAA's precipitation reconstruction dataset (PRE) \cite{NOAA-precip}. It provides monthly global precipitation constructed on a 2.5-degree latitude $\times$ 2.5-degree longitude grid over the global region (72 $\times$ 144) for the period from Jan. 1948 to Aug. 2018 \cite{chen2002global}. Following \cite{manohar2018data}, the first 16 years’ SST data are selected as the training set, and the remaining data are used as the test set. As for the PRE dataset, the first 70\% of spatiotemporal field snapshots are chosen as the training set, and the remaining 30\% snapshots are used as the test set for performance evaluation.

\subsubsection{Evaluation Metrics}
The proposed framework is evaluated by using two metrics, MSE@$N$, and VAR@$N$, where $N$ is the number of sensor nodes of the WSN. The optimal values of $N$ for the SST dataset and the PRE dataset are obtained according to \cite{manohar2018data}, which are 302 and 206, respectively. MSE@$N$ is the Mean Square Error that measures the average of the squared error between the predictions and the ground truth. It is calculated as: 
\begin{equation}
\textrm{MSE@}N = \frac{1}{m\cdot M}\sum_{i=1}^m\sum_{j=1}^M(\boldsymbol{\Phi}_{ij}-\tilde{\boldsymbol{\Phi}}_{ij})^2.
\end{equation}

VAR@$N$ calculates the mean variance of MSE for all reconstruction. It evaluates the variance of the long-term prediction. Specifically, a low VAR@$N$ indicates that the MSE of the predicting test snapshots will not change significantly in most locations. Hence, the model can perform a more generalized reconstruction and has a greater prediction ability. 
\begin{equation}
\textrm{VAR@}N = \frac{1}{m}\sum_{i=1}^m(\frac{1}{M}\sum_{j=1}^M(\boldsymbol{\Phi}_{ij}-\tilde{\boldsymbol{\Phi}}_{ij})^2-\textrm{MSE@}N)^2.
\end{equation}

Thus, the aim of the simulation is to test if the proposed method has both low MSE@$N$ value and low VAR@$N$ value.

\subsubsection{Benchmark Algorithms}
The proposed algorithm, is compared with the following benchmark algorithms:  
\begin{itemize}
\item \textbf{Q-DEIM}: QR factorization with the discrete empirical interpolation method (Q-DEIM) utilizes a greedy approximation solution provided by the matrix QR factorization with column pivoting. The sampling scheme of Q-DEIM seeks rows of $\boldsymbol{\Psi}_r$, corresponding to the deployment locations of the WSN in the spatiotemporal field that will optimally condition the inversion of the measurement matrix \cite{drmac2016new, manohar2018data}. The measurement matrix $\mathbf{C}$ is then extracted from the corresponding sampling locations of $\boldsymbol{\Psi}_r$.
\item \textbf{CS}: Compressive sensing (CS) can efficiently compress and reconstruct a signal by solving suitable underdetermined linear systems  \cite{tipping2001sparse, candes2008introduction, zhang2014spatiotemporal}. Similar to Q-DEIM, the signal is compressed via the measurement matrix: $\mathbf{y} = \mathbf{C}\cdot \boldsymbol{\phi}$. 
\item \textbf{RAND}: RAND is a baseline algorithm in which the sampling locations are randomly deployed in the field. The reconstruction method is similar to Q-DEIM, except for using random sampling locations. The random seed for generating the sampling locations is the same as that used in CS.  
\item \textbf{VAE}: Variational autoencoder (VAE) learns a low-dimensional representation for a high-diemsnional signal, which can reduce the dimension significantly \cite{kingma2013auto}. However, VAE encodes the signal into a low-dimensional latent domain, where in situ measurements are unavaliable. Therefore, in the present simulation, the VAE modified to learn to reconstruct signals from randomly sampled data.
\item \textbf{RAND-DL}: Random sampling-based DL model is a baseline method for the proposed model. It replaces the sampling optimization scheme in the proposed model with random sampling to test the effectiveness of the sampling optimization scheme. The random sampling scheme is generated with the same seed as in VAE.
\end{itemize}

\subsection{Results and Performance Comparison}
Table \ref{table:GEN-statistic} summarizes the performance of the SST and the PRE datasets based on the evaluation metrics described in Section \ref{subsection:GEN-Experiment-setup}. The results in this table were obtained as the average value of 10 trials. The boldface values are the best results achieved by all algorithms, while the underlined values are the best results obtained by the benchmark algirhtms. Note that the performance improvement is calculated based on the best bencchmark algorithm:
\begin{equation}
\frac{\textrm{benchmark - the proposed method}}{\textrm{benchmark}}\times 100\%
\end{equation}

\begin{table*}[ht]
    \centering
    \setlength\aboverulesep{0pt}\setlength\belowrulesep{0pt}
    \setcellgapes{3pt}\makegapedcells
    \caption{Comparison of performance using the two datasets.}\label{table:GEN-statistic}
    \begin{tabular}{|c|c|c|c|c|c|c|c|c|}
    \toprule
    {}  & {Metric} & QDEIM  & CS& RAND & VAE&RAND-DL& The proposed method & Improv. \\ 
    \midrule\midrule
    \multirow{8}{*}{SST} 
    & MSE@100 & 1.783 & 2.083 & 25.520  & \underline{0.4192} & 0.4499 & \textbf{0.3910} & 6.73\% \\
 & MSE@200 & 2.329 & 1.564 & 27.477  & \underline{0.4056} & 0.4410 & \textbf{0.3552} & 12.43\% \\
 & MSE@302 & 2.534 & 1.400 & 26.748  & \underline{0.4071} & 0.4638 & \textbf{0.3326} & 18.30\% \\
 & MSE@400 & 2.530 & 1.253 & 27.904  & \underline{0.4153} & 0.4599 & \textbf{0.3215} & 22.59\% \\
 & VAR@100 & 0.527 & 1.174 & 163.120 & \underline{0.0087} & 0.0120 & \textbf{0.0079} & 9.20\% \\
 & VAR@200 & 0.654 & 0.426 & 195.780 & \underline{0.0080} & 0.0104 & \textbf{0.0061} & 23.75\% \\
 & VAR@302 & 1.049 & 0.250 & 155.660 & \underline{0.0096} & 0.0121 & \textbf{0.0056} & 41.67\% \\
 & VAR@400 & 1.076 & 0.192 & 124.730 & \underline{0.0098} & 0.0109 & \textbf{0.0051} & 47.96\%\\
    \midrule\midrule
    \multirow{8}{*}{PRE}
    & MSE@50  & 0.857 & 1.216 & 19.213  & 0.4926 & \underline{0.4636} & \textbf{0.2848} & 38.57\% \\
 & MSE@100 & 0.981 & 1.013 & 23.400  & \underline{0.3202} & 0.3656 & \textbf{0.2558} & 20.11\% \\
 & MSE@206 & 1.147 & 1.019 & 30.260  & \underline{0.2773} & 0.2793 & \textbf{0.2396} & 13.60\% \\
 & MSE@300 & 1.424 & 0.900 & 33.822  & \underline{0.2526} & 0.2556 & \textbf{0.2350} & 06.97\% \\
 & VAR@50  & 0.193 & 0.550 & 185.360 & 0.0392 & \underline{0.0358} & \textbf{0.0074} & 79.33\% \\
 & VAR@100 & 0.406 & 0.338 & 137.070 & \underline{0.0119} & 0.0148 & \textbf{0.0056} & 52.94\% \\
 & VAR@206 & 0.440 & 0.342 & 106.640 & \underline{0.0069} & 0.0070 & \textbf{0.0050 }& 27.54\% \\
 & VAR@300 & 0.528 & 0.227 & 81.895  & \underline{0.0052} & \underline{0.0052} & \textbf{0.0048} & 7.69\% \\
    \bottomrule
    \end{tabular}
\end{table*}

In general, the proposed method achieves the best reconstruction accuracy for long-term spatiotemporal field reconstruction and prediction in both datasets, given a limited set of observations. The minimum reconstruction errors of the proposed method in the SST and the PRE datasets are 0.3215 and 0.2350, respectively. It is seen that the proposed method improves the reconstruction accuracy significantly, which is at least 6.73\% higher than the results generated by the benchmark algorithms. The model robustness in predicting long-term spatiotemporal field is also improved dramatically. For all tests cases, the proposed method obtains both the lowest reconstruction error and the lowest variance in predicting long-term precipitation data. Therefore, it can be concluded that the proposed method can produce a more generalized and robust result in reconstructing and predicting a complex spatiotemporal signal.

VAE and RAND-DL generate the second-best results in reconstructing and predicting the spatiotemporal data, given a few observations. VAE outperfoms RAND-DL in almost all test cases. Althrough RAND-DL outperforms VAE in the deployment of 50 sensor using the PRE dataset, the improvement is very limited. The results generated by Q-DEIM and CS is worse than the ones generated by VAE and RAND-DL. Q-DEIM can achieve better performance if the number of sensor nodes is small since its mechanism attempts to extract the essential information from the SVD basis. However, as the number of sensor nodes increases, CS gradually achieves better results. RAND produces the worst results with both datasets, as it generates the highest reconstruction error and variance.

\begin{figure*}[h]
    \centering
    \begin{subfigure}[b]{0.225\textwidth}
        \includegraphics[width=\textwidth]{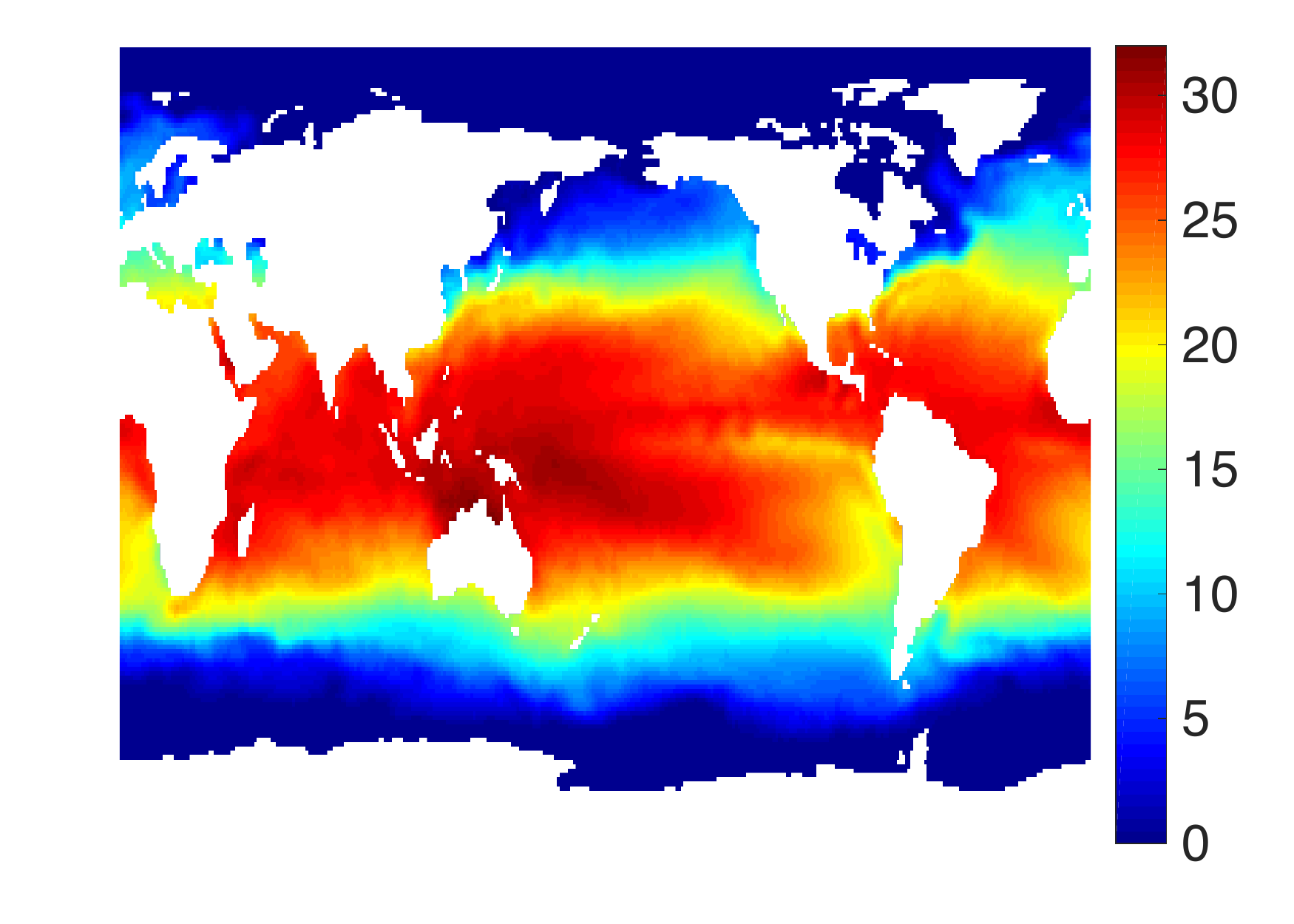}
        \caption{Ground truth}
    \end{subfigure}
    \begin{subfigure}[b]{0.225\textwidth}
        \includegraphics[width=\textwidth]{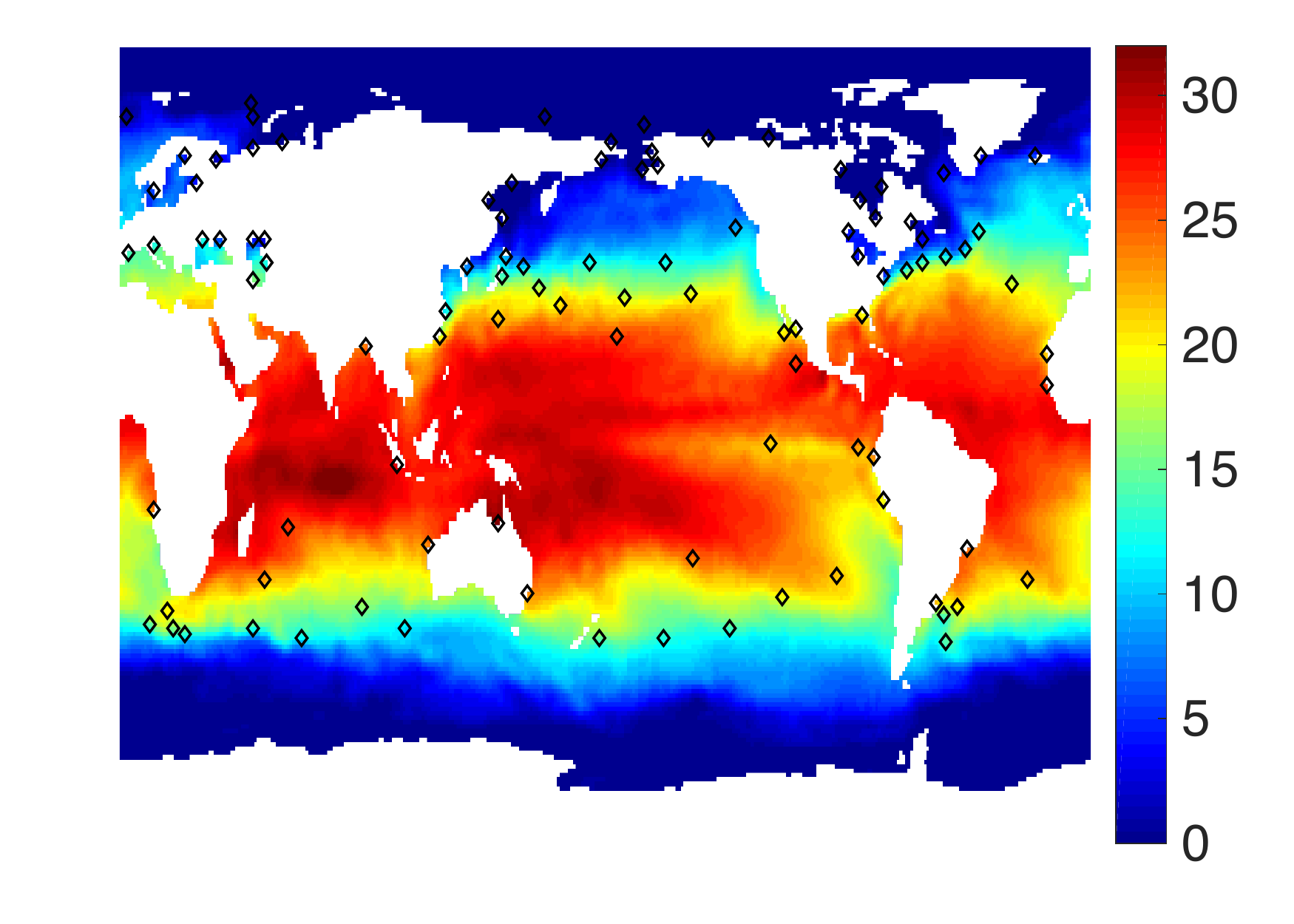}
        \caption{Q-DEIM}
    \end{subfigure}
    \begin{subfigure}[b]{0.225\textwidth}
        \includegraphics[width=\textwidth]{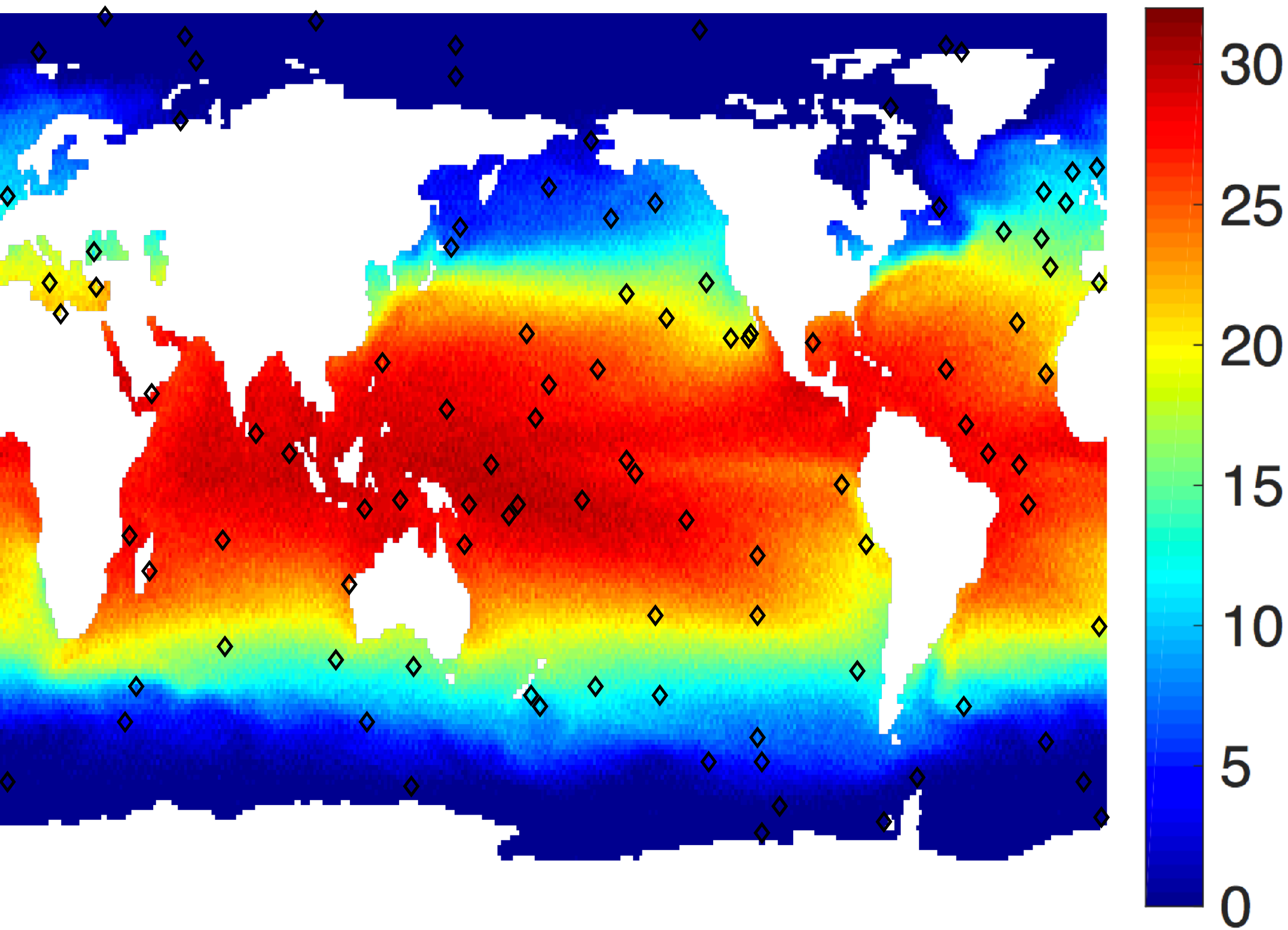}
        \caption{CS}
    \end{subfigure}
    \begin{subfigure}[b]{0.225\textwidth}
        \includegraphics[width=\textwidth]{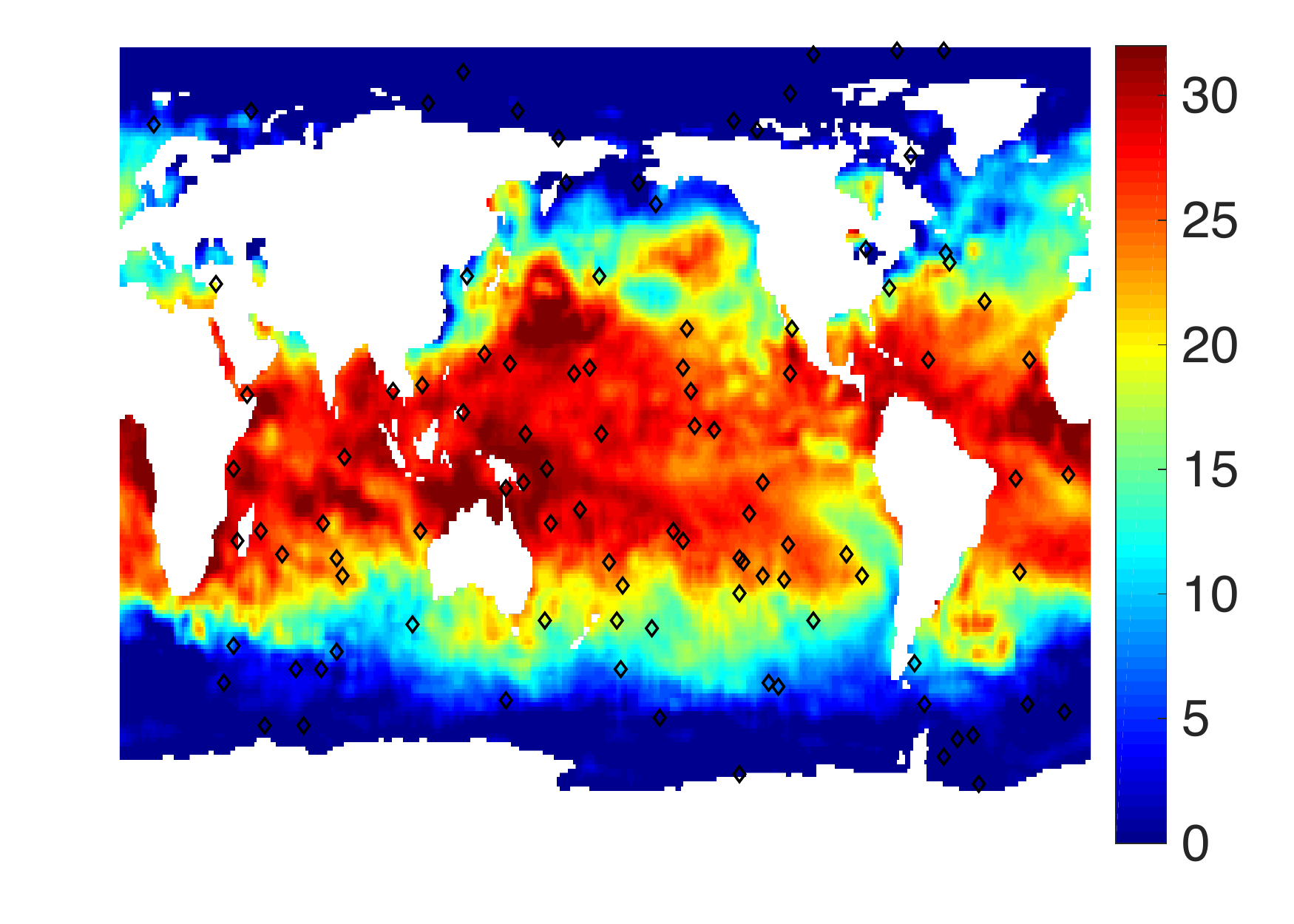}
        \caption{RAND}
    \end{subfigure}
    \begin{subfigure}[b]{0.225\textwidth}
        \includegraphics[width=\textwidth]{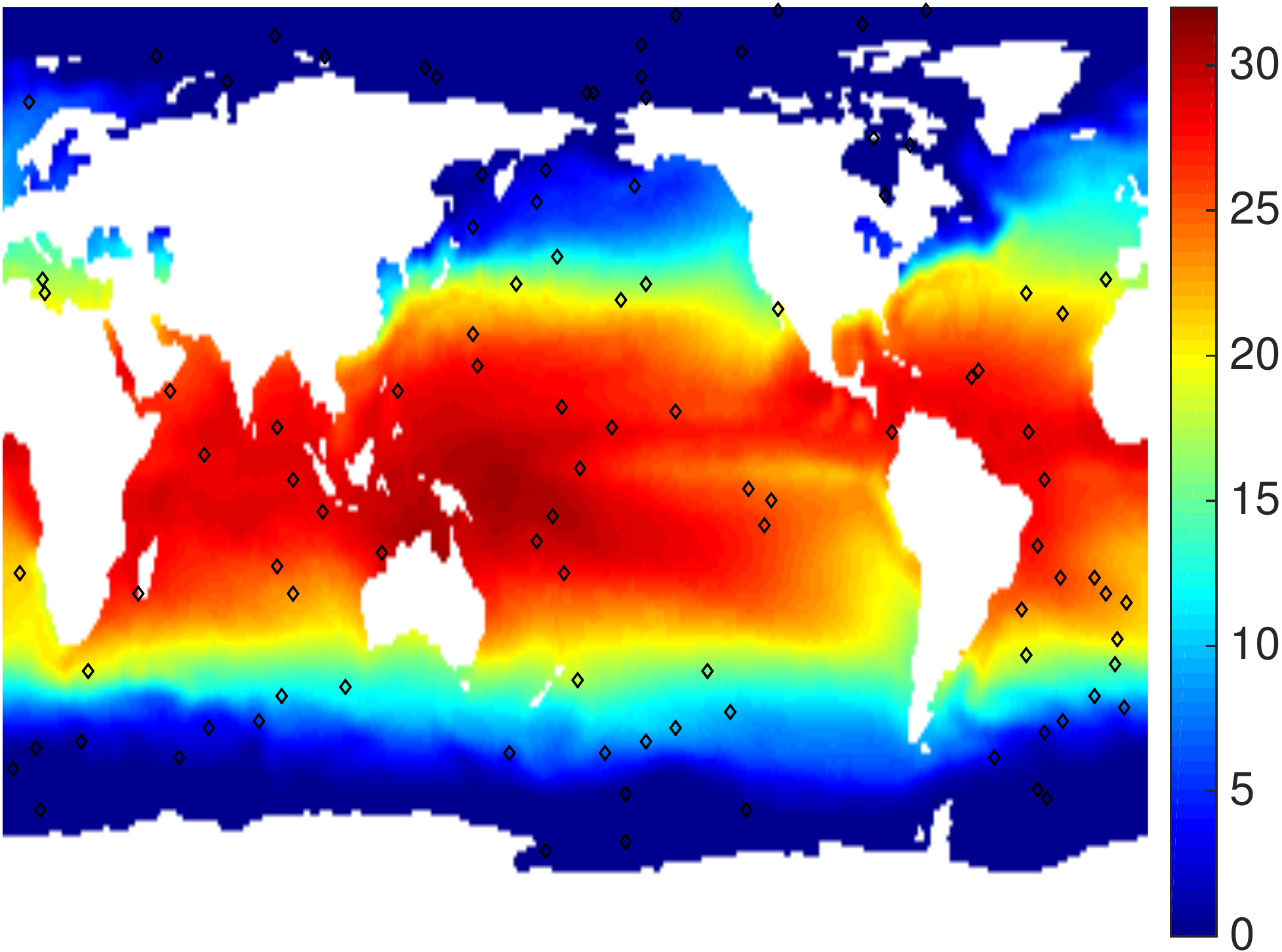}
        \caption{VAE}
    \end{subfigure}
    \begin{subfigure}[b]{0.225\textwidth}
        \includegraphics[width=\textwidth]{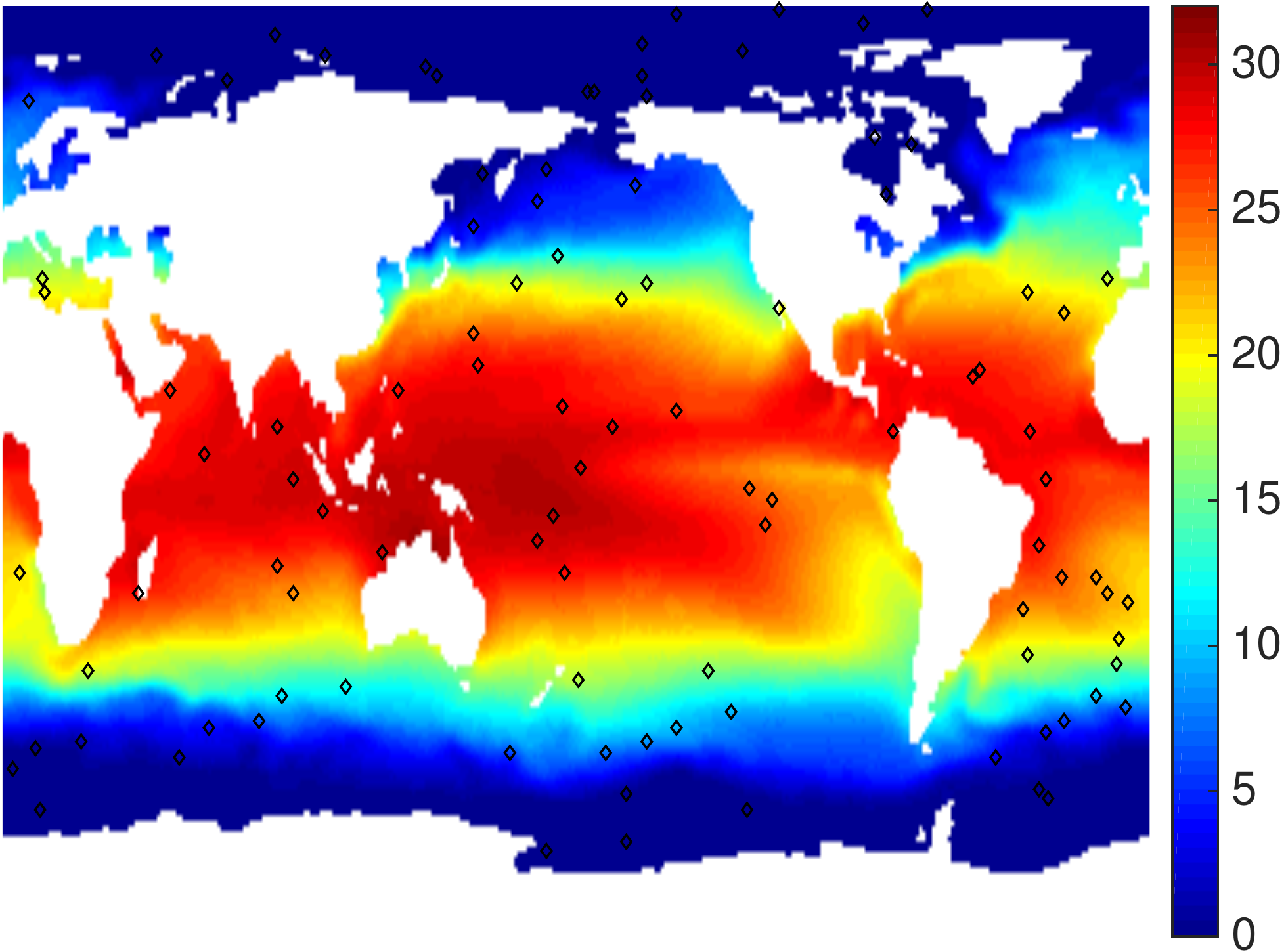}
        \caption{RAND-DL}
    \end{subfigure}    
    \begin{subfigure}[b]{0.225\textwidth}
        \includegraphics[width=\textwidth]{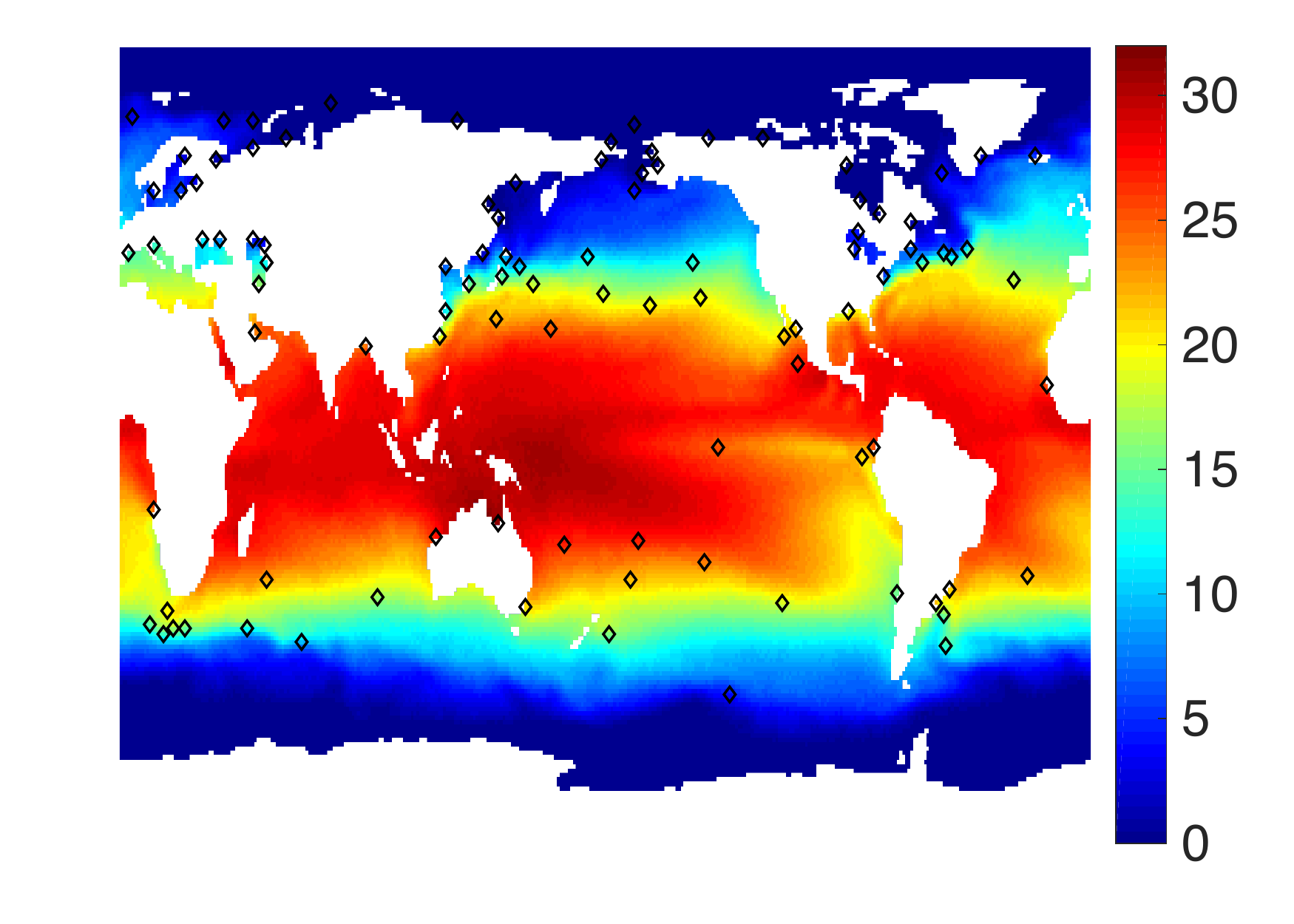}
        \caption{The proposed method}
    \end{subfigure}
    \caption{Ground truth of the tested spatiotemporal field and reconstruction from the proposed and the benchmark algorithms using the SST dataset (Black diamonds: sampling locations.)}\label{GEN-SST-RECO}
\end{figure*}

Figure \ref{GEN-SST-RECO} presents one snapshot of reconstruction results for the SST dataset with 100 sensor nodes, using the proposed and the benchmark algorithms. The black diamonds in this figure indicate the sensor deployment locations. The proposed method produces the visibly best reconstruction when compared with the ground truth. Details of the ground truth are fully reconstructed, and critical features of global SST, such as ¬¬current and El Ni\~{n}o, are captured. As shown in Figure \ref{GEN-SST-RECO} (g), the proposed method tends to deploy sensor nodes near the boundaries of different temperature levels.

VAE and RAND-DL also produces visibly good reconstruction results. Q-DEIM produces slightly worse reconstruction results than the DL-based methods. The reconstructed SST increases significantly over the Indian Ocean. Nevertheless, the SST reconstruction for other oceans is successful. Note that the sensor deployment locations selected by Q-DEIM are close to the borders of the continents, which may occur due to higher dynamics in coastal areas. However, although the information collected in coastal regions has a higher variance than in blue water, it is difficult to generalize local information to a global situation. Instead, more observations in the oceanic area will help to produce a better reconstruction model. In comparison, the proposed method learns more information by collecting in situ measurements in a more distributed manner over the spatiotemporal field.

CS generates less accurate reconstruction results than Q-DEIM. The temperature distributions along the latitude are captured; however, the details of the ground truth are lost, which blurs the reconstructed field image. Information about inland bodies of water, such as the Mediterranean Sea, the Black Sea, and the Great Lakes, is also lost. Moreover, as CS-based approaches do not provide a scheme to optimize the sensor deployment locations, the measurement matrix is randomly generated. Therefore, the reconstruction process may have higher variance if the selected sampling locations are ineffective.

RAND produces the worst reconstruction results. The reconstructed SST is significantly higher than the ground truth in several oceans. As in CS, the sampling locations are randomly selected; therefore, less valid observations are obtained. As a result, RAND cannot provide satisfactory results in reconstruction and estimation.

\begin{figure*}[t]
    \centering
    \begin{subfigure}[b]{0.225\textwidth}
        \includegraphics[width=\textwidth]{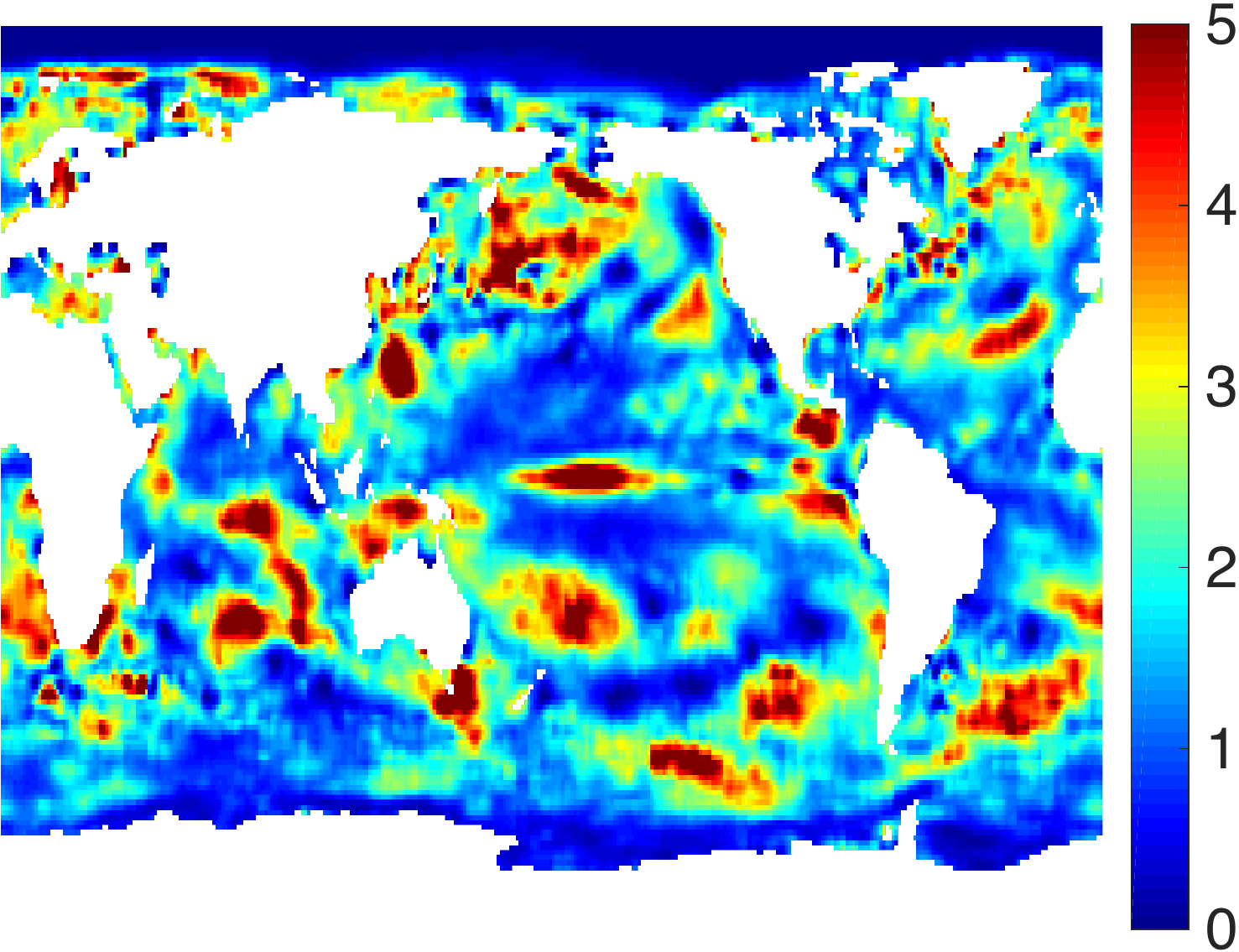}
        \caption{Q-DEIM}
    \end{subfigure}
    \begin{subfigure}[b]{0.225\textwidth}
        \includegraphics[width=\textwidth]{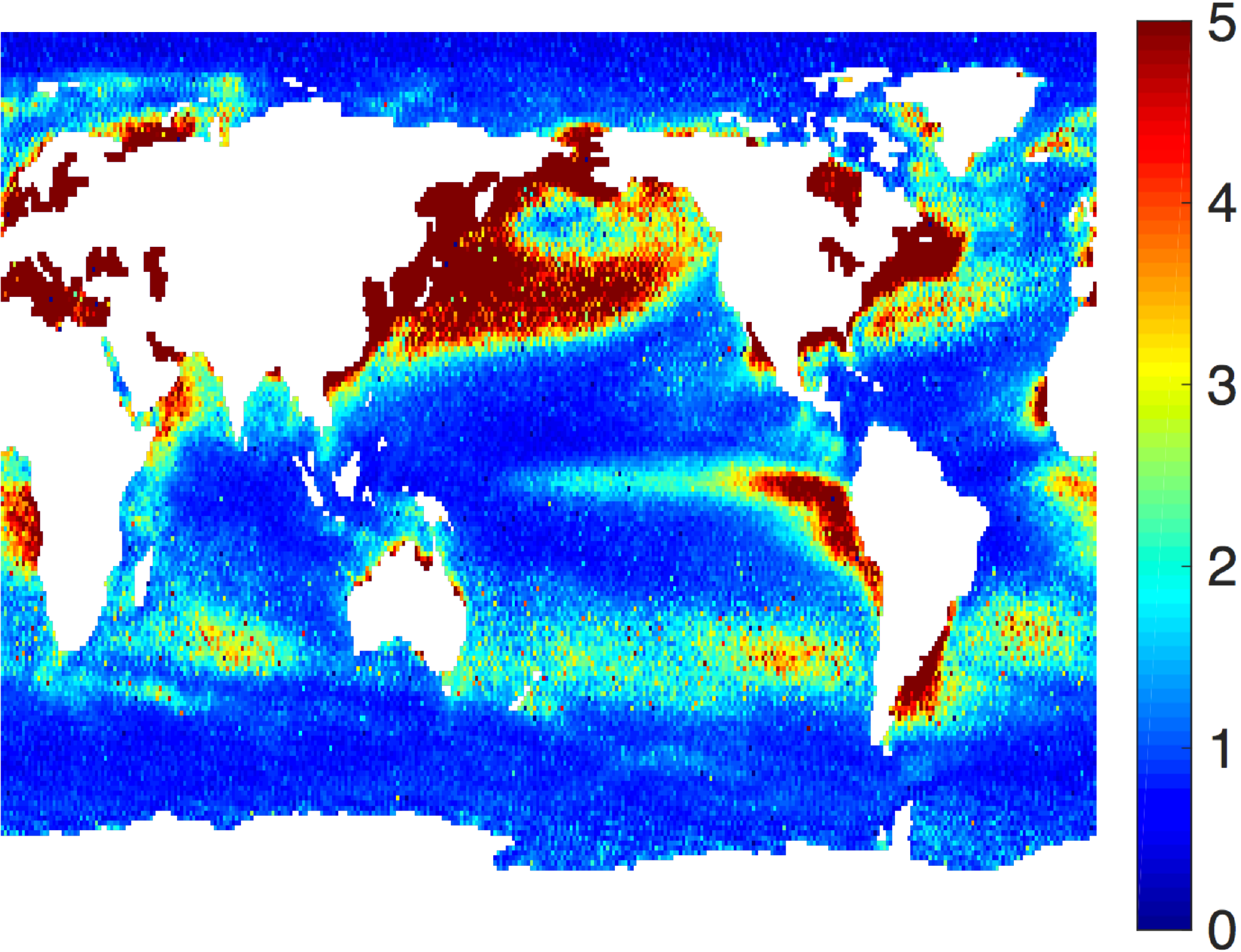}
        \caption{CS}
    \end{subfigure}
    \begin{subfigure}[b]{0.215\textwidth}
        \includegraphics[width=\textwidth]{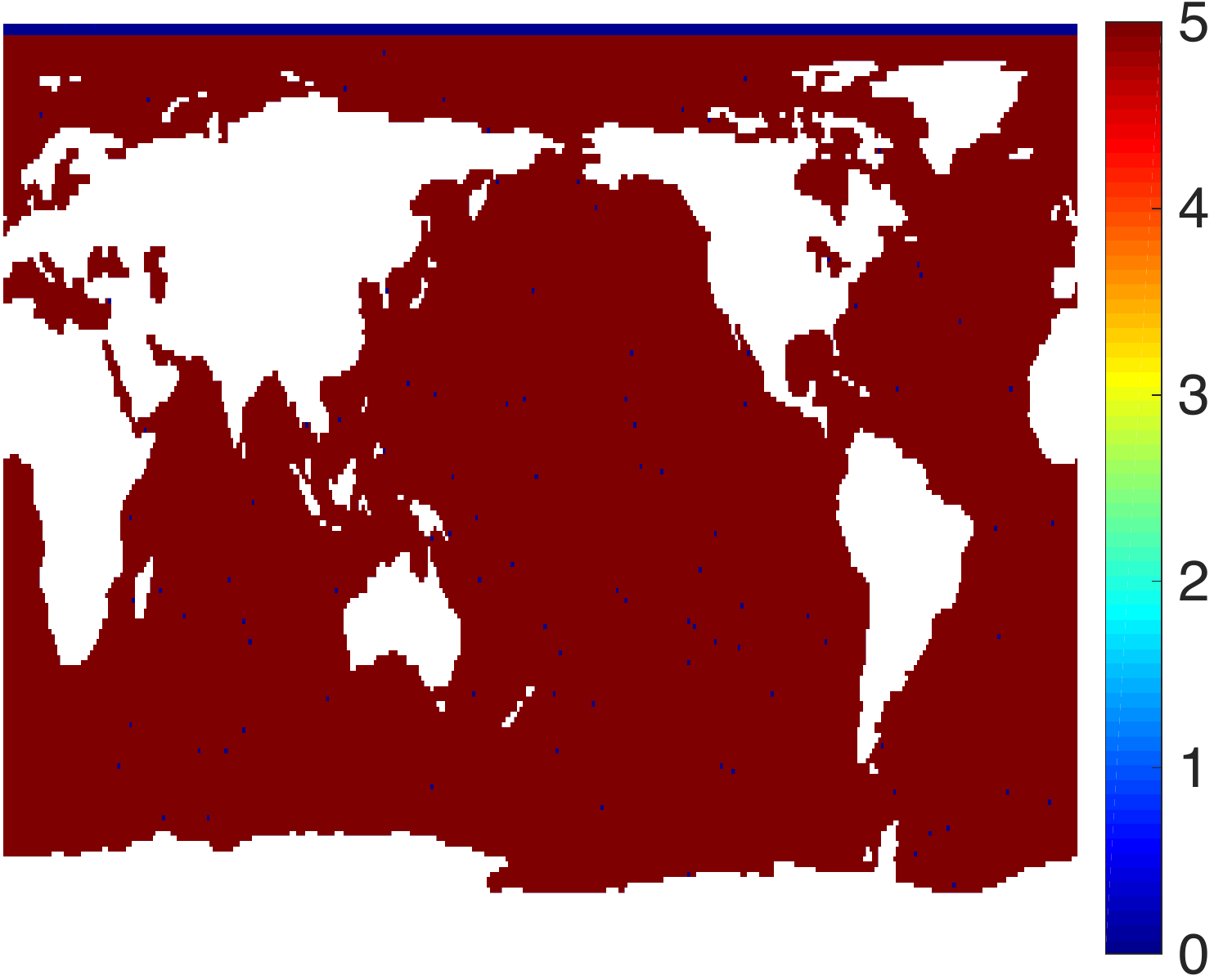}
        \caption{RAND}
    \end{subfigure}
    
    \begin{subfigure}[b]{0.225\textwidth}
        \includegraphics[width=\textwidth]{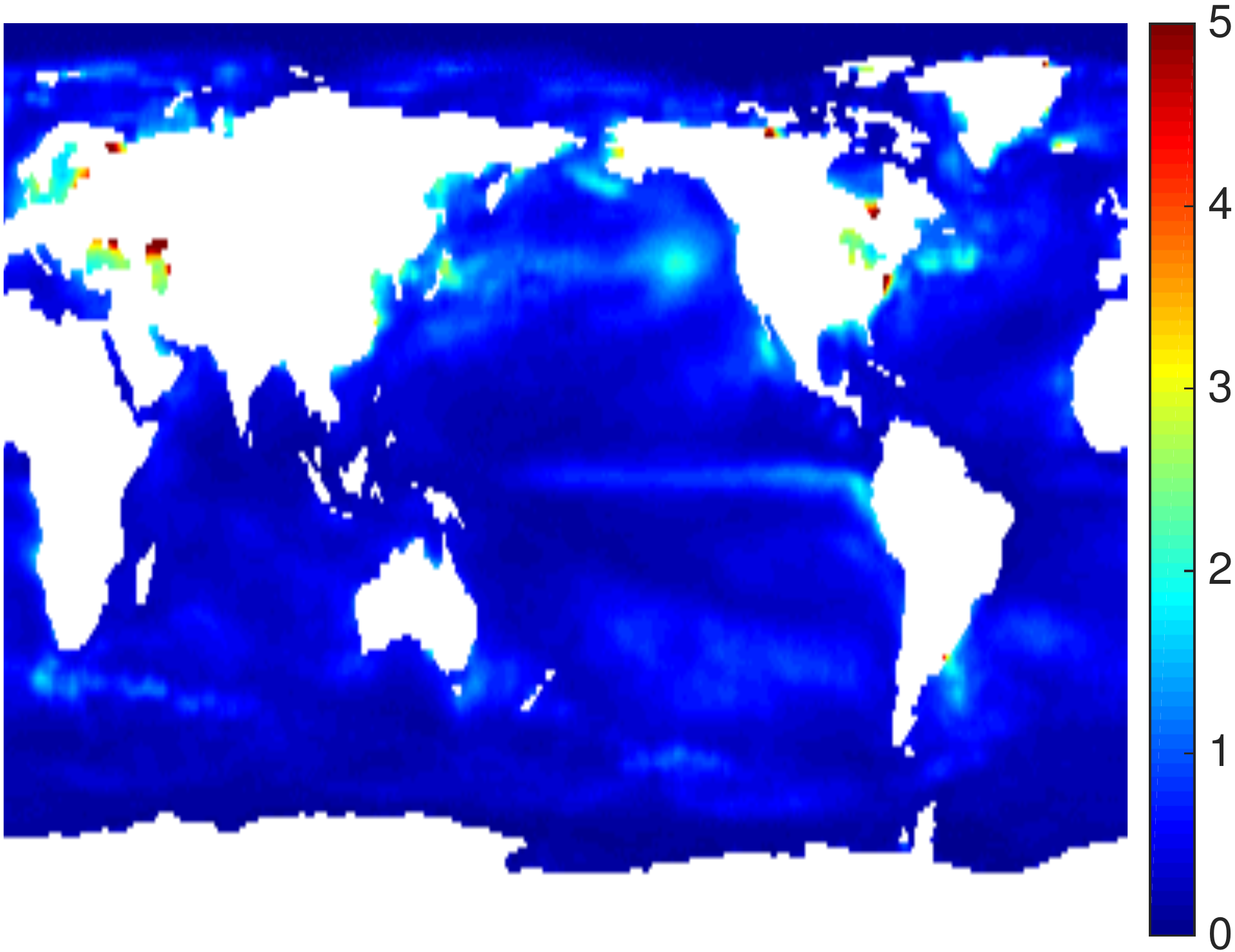}
        \caption{VAE}
    \end{subfigure}
    \begin{subfigure}[b]{0.225\textwidth}
        \includegraphics[width=\textwidth]{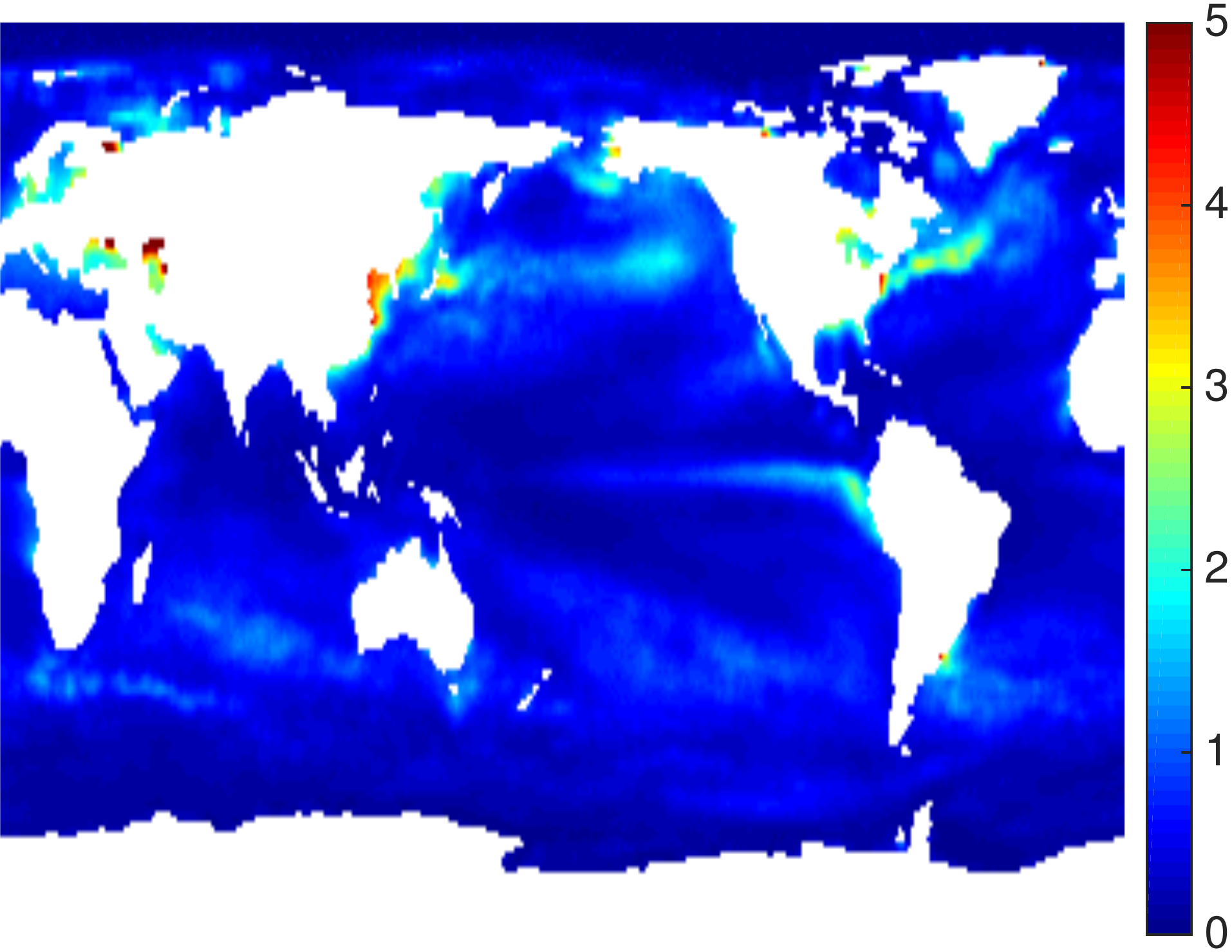}
        \caption{RAND-DL}
    \end{subfigure}
    \begin{subfigure}[b]{0.225\textwidth}
        \includegraphics[width=\textwidth]{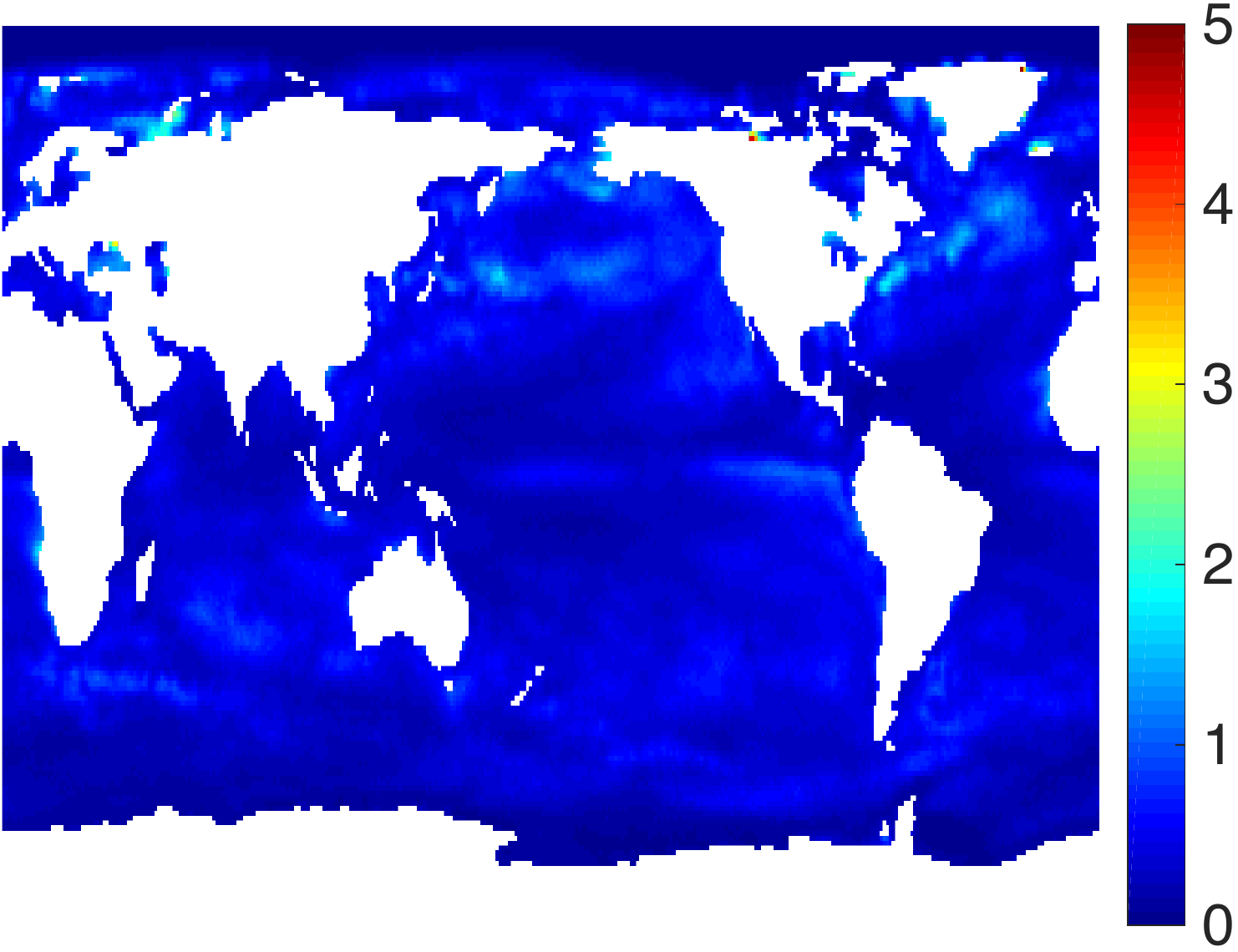}
        \caption{The proposed method}
    \end{subfigure}
    \caption{Reconstruction variance for all test snapshots in the SST dataset.}\label{GEN-SST-RECO-VAR}
\end{figure*}

Figure \ref{GEN-SST-RECO-VAR} presents the prediction variance for all test sets, which examines the reconstruction variance for long-term prediction. The presented variance map is calculated pixel-wise and evaluates the significance of the MSE variance for all future predictions. Hence, a higher variance shown in this figure indicates a lower confidence for future prediction.

As shown in Figure \ref{GEN-SST-RECO-VAR} (f), the proposed method achieves the lowest variance across the entire field. The variance for long-term prediction only occurs in some offshore areas. For example, the variance in the northern Pacific Ocean is high, which may be affected by the Aleutian Islands for the SST disturbance introduced by the archipelago. VAE and RAND-DL also achieves similar results, but the resulting reconstruction variance elevates in the mentioned offshore areas. The results for Q-DEIM and CS are presented in Figure  \ref{GEN-SST-RECO-VAR} (a) - (b), which have higher variance in predicting future spatiotemporal fields than the DL-based methods. The high prediction variance also indicates that the selected sampling locations cannot provide sufficient information for the reconstruction model. As a result, when the environment changes, the Q-DEIM and CS cannot perform well. Thus, their reconstruction and prediction models are not generalized and are not robust for estimating environmental changes. The result for RAND is illustrated in Figure  \ref{GEN-SST-RECO-VAR} (c), which produces the worst result. The result show that the long-term prediction variance of RAND exceeds five in almost the entire area. Hence, RAND has the highest variance due to randomly deployed sensor nodes. It follows that the proposed method outperforms CS, Q-DEIM, and RAND significantly for optimizing sensor deployment locations in global SST reconstruction.

\begin{figure*}[t]
     \centering
     \begin{subfigure}[b]{0.32\textwidth}
         \includegraphics[width=\textwidth]{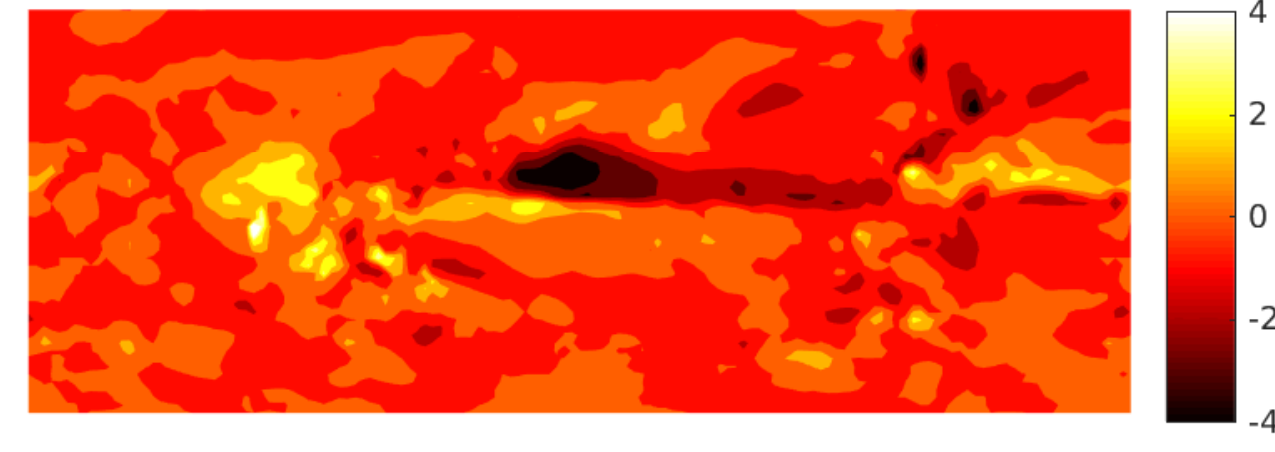}
         \caption{Ground truth}
     \end{subfigure}
     \begin{subfigure}[b]{0.32\textwidth}
         \includegraphics[width=\textwidth]{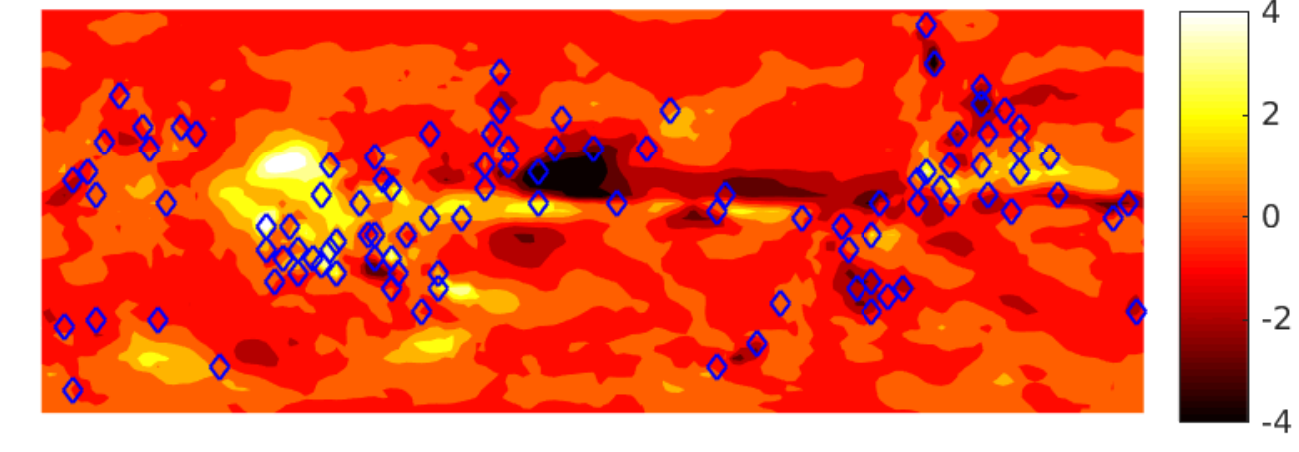}
         \caption{Q-DEIM}
     \end{subfigure}
     \begin{subfigure}[b]{0.31\textwidth}
         \includegraphics[width=\textwidth]{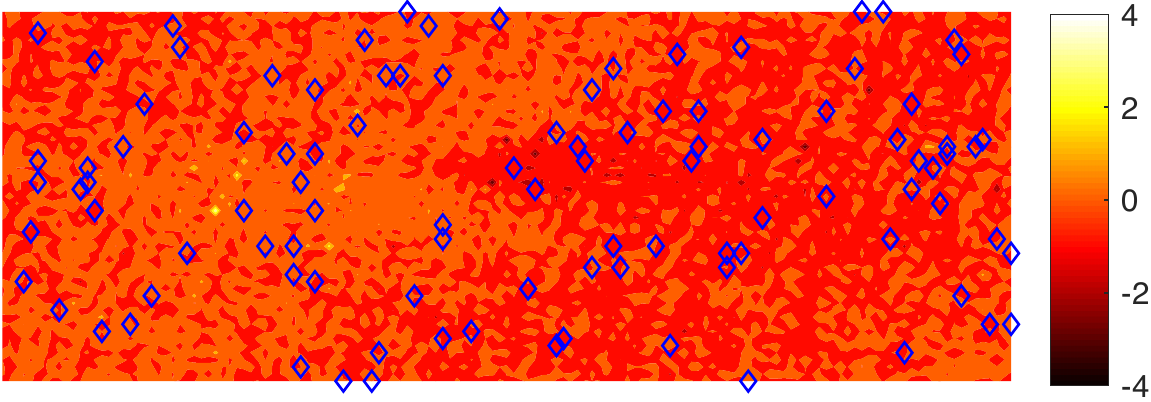}
         \caption{CS}
     \end{subfigure}
     \begin{subfigure}[b]{0.32\textwidth}
         \includegraphics[width=\textwidth]{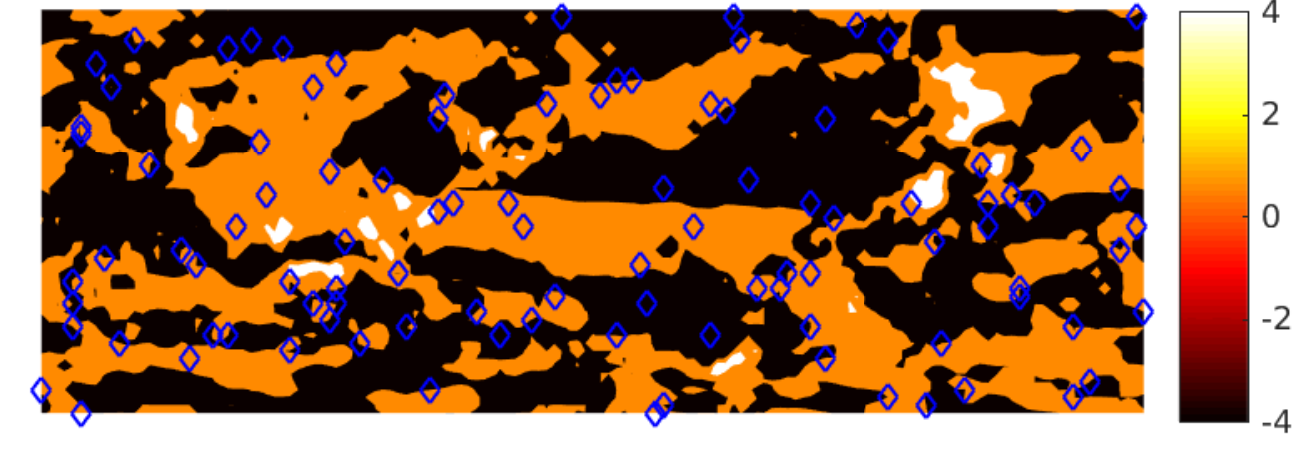}
         \caption{RAND}
     \end{subfigure}
     \begin{subfigure}[b]{0.32\textwidth}
         \includegraphics[width=\textwidth]{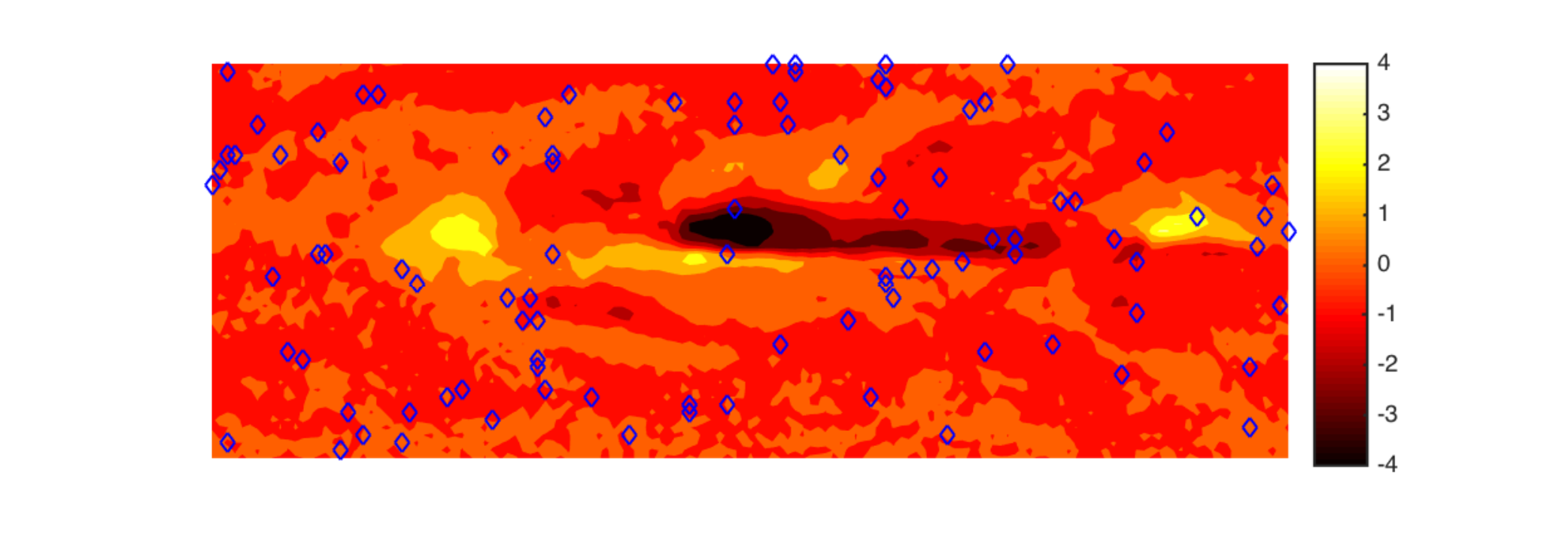}
         \caption{VAE}
     \end{subfigure}
     \begin{subfigure}[b]{0.32\textwidth}
         \includegraphics[width=\textwidth]{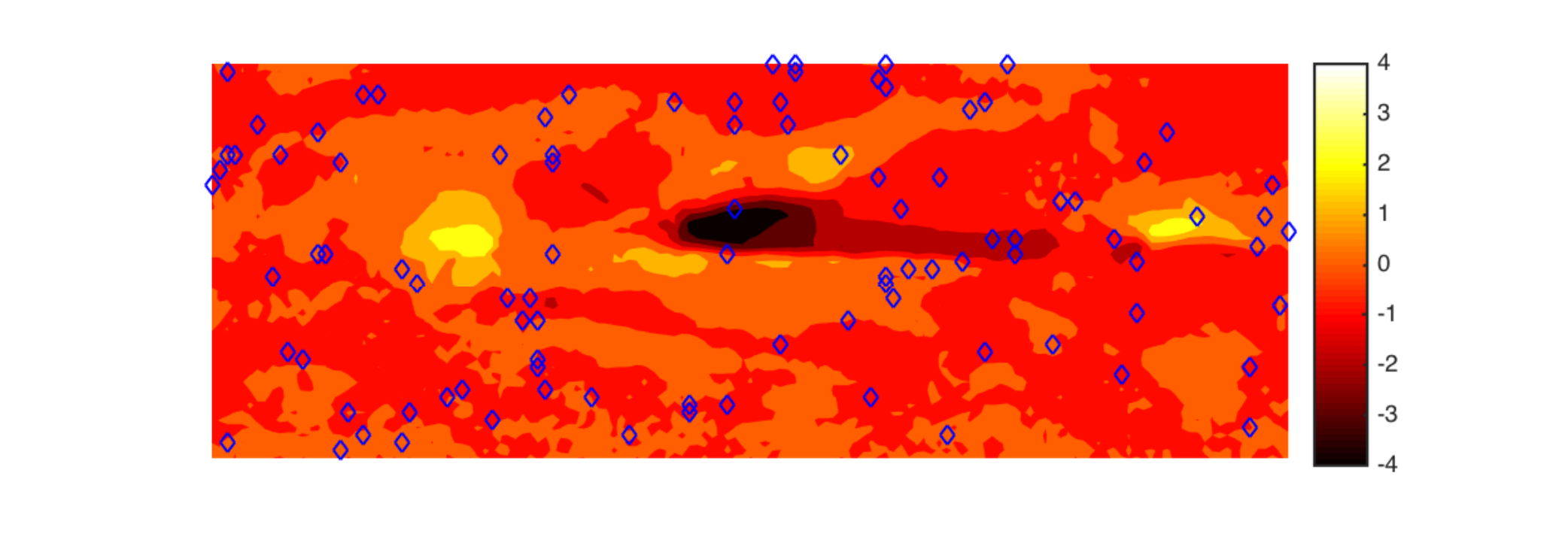}
         \caption{RAND-DL}
     \end{subfigure}
     \begin{subfigure}[b]{0.32\textwidth}
         \includegraphics[width=\textwidth]{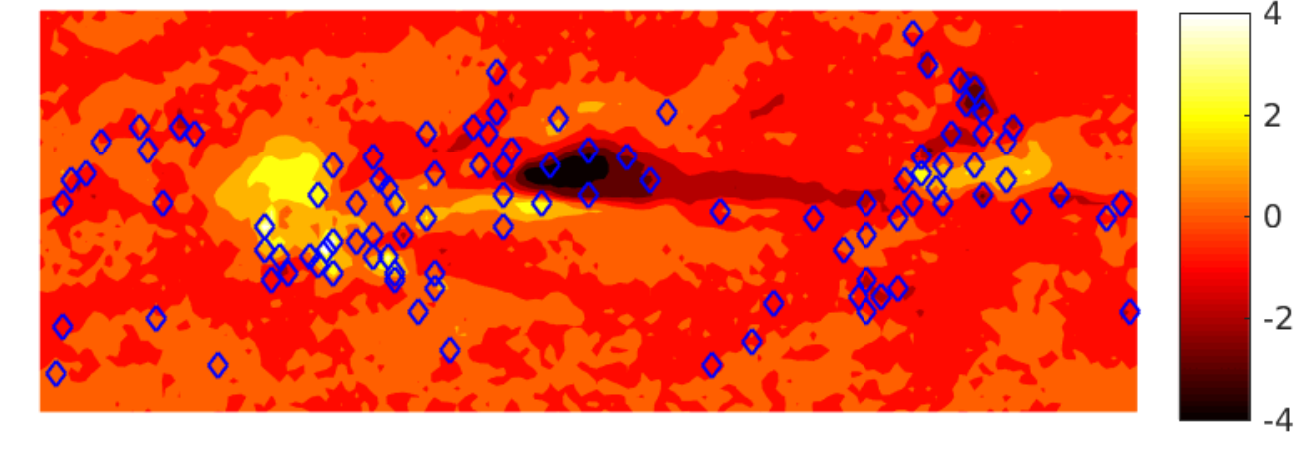}
         \caption{The proposed method}
     \end{subfigure}
     \caption{Ground truth of the testing spatiotemporal field and reconstruction from the proposed and benchmark algorithms in the PRE dataset. (Blue diamonds: sampling locations.)}\label{GEN-PREC-RECO}
\end{figure*}

Figure \ref{GEN-PREC-RECO} presents the reconstruction results for the PRE dataset. Similar to the results for the SST dataset, the proposed method produces the best reconstruction results for the PRE dataset, which followed by VAE and RAND-DL. Q-DEIM produces a worse reconstruction result, which is less generalized and loses some feature. Overall, the spatiotemporal field is accurately reconstructed by the DL-based methods. In contrast, Q-DEIM tends to over-fit the training data. Hence, slight changes in precipitation may have a significant impact on the reconstruction, causing a higher reconstruction MSE.

As shown in Figure \ref{GEN-PREC-RECO} (c)-(d), CS and RAND cannot produce valid reconstruction for the PRE dataset. Significant disturbance and inaccuracy are introduced, and the original spatiotemporal field is poorly reconstructed.

\begin{figure*}
   \centering
    \begin{subfigure}[b]{0.32\textwidth}
       \includegraphics[width=\textwidth]{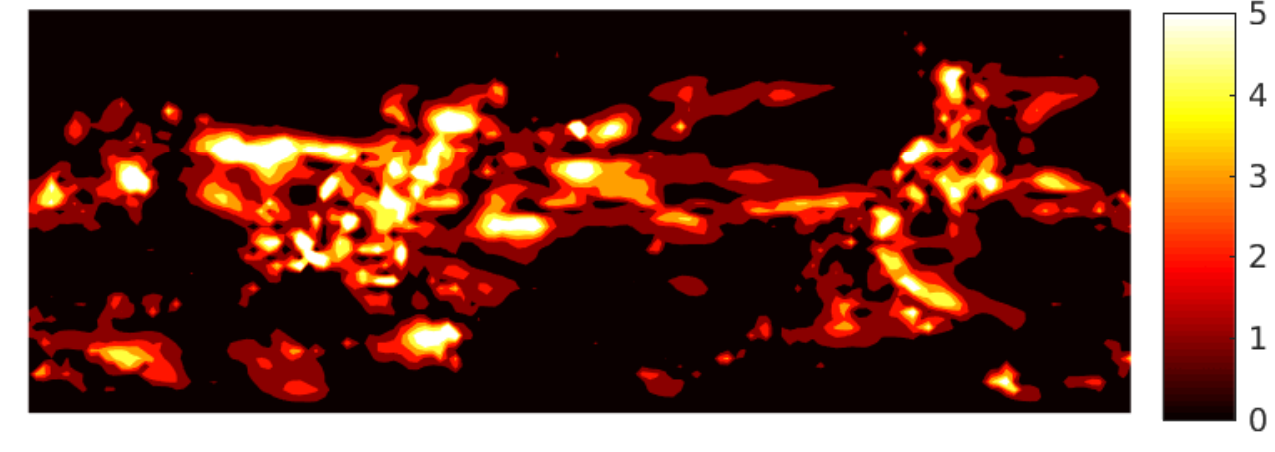} 
       \caption{Q-DEIM}
    \end{subfigure}
    \begin{subfigure}[b]{0.32\textwidth}
       \includegraphics[width=\textwidth]{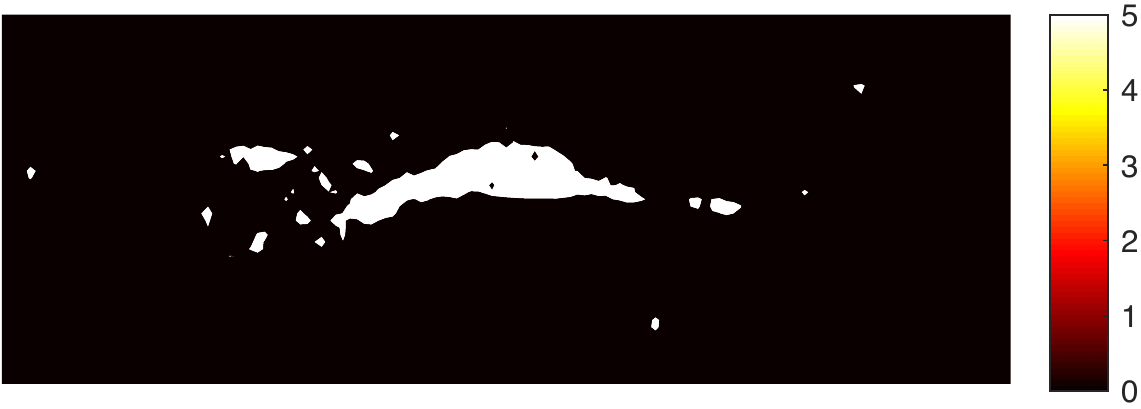}
       \caption{CS}
    \end{subfigure} 
    \begin{subfigure}[b]{0.32\textwidth}
        \includegraphics[width=\textwidth]{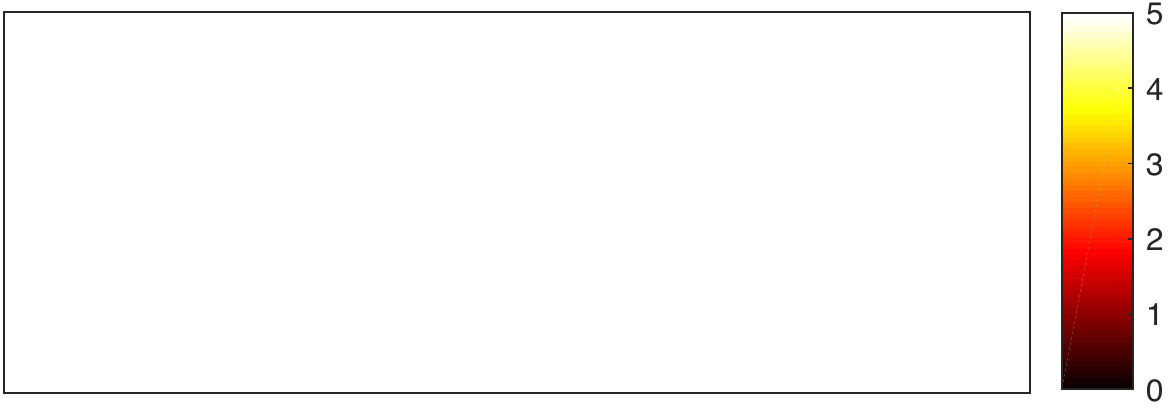}
        \caption{RAND}
    \end{subfigure}
    \begin{subfigure}[b]{0.32\textwidth}
       \includegraphics[width=\textwidth]{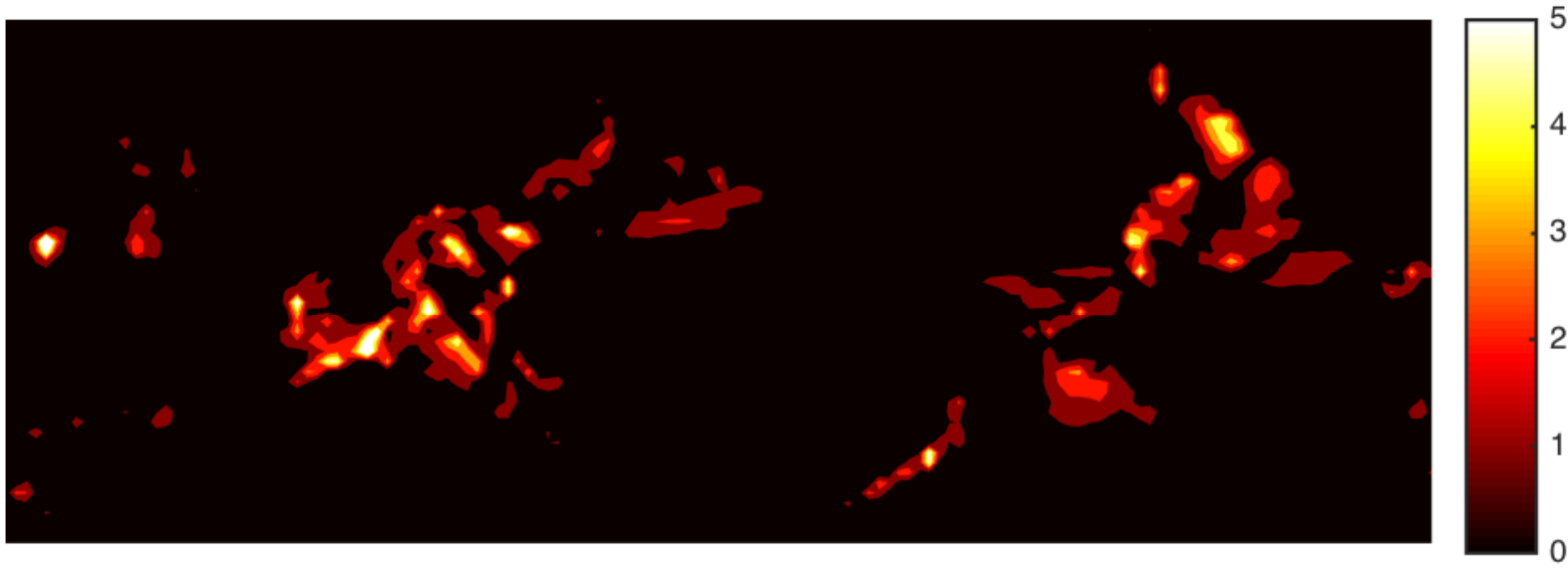}
       \caption{VAE}
    \end{subfigure}
    \begin{subfigure}[b]{0.32\textwidth}
       \includegraphics[width=\textwidth]{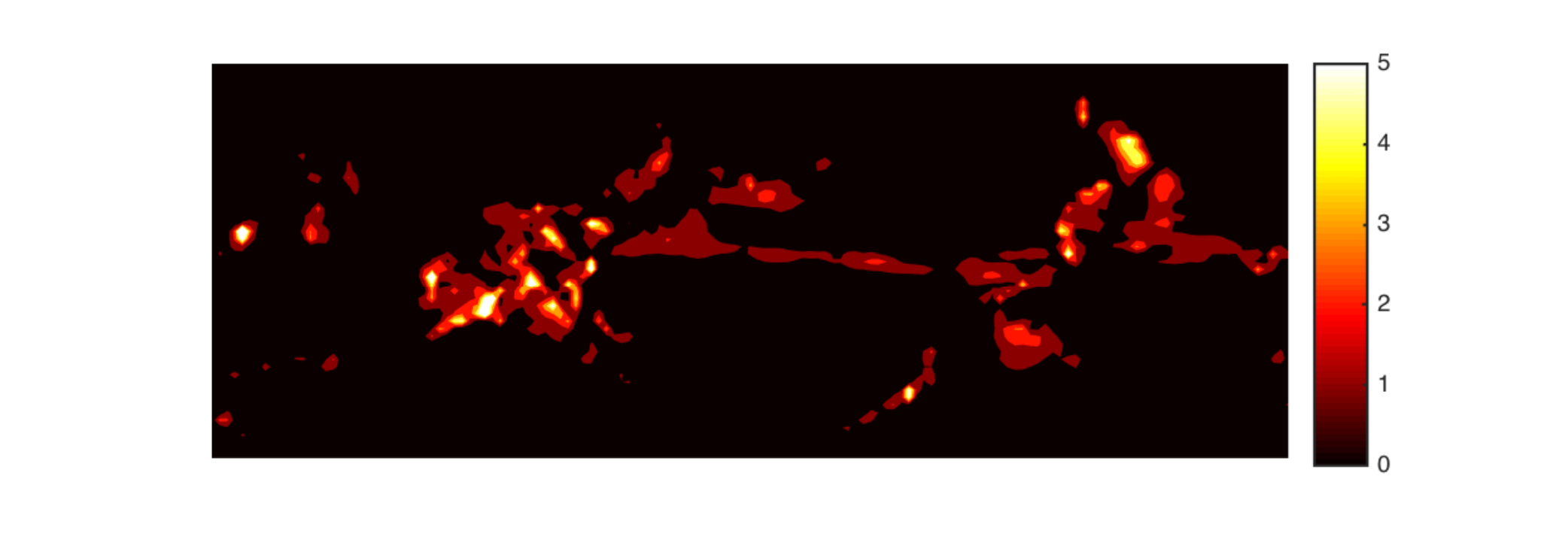}
       \caption{RAND-DL}
    \end{subfigure}
    \begin{subfigure}[b]{0.32\textwidth}
       \includegraphics[width=\textwidth]{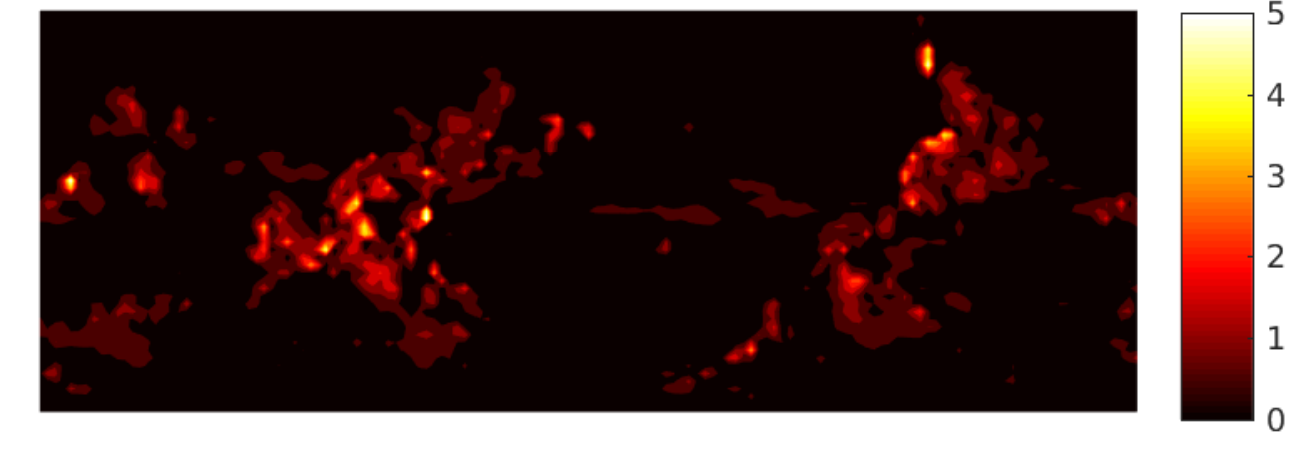}
       \caption{The proposed method}
    \end{subfigure}
    \caption{Reconstruction variance for all test snapshots in the PRE dataset.}\label{GEN-PREC-RECO-VAR}
\end{figure*}

Similar results are obtained from the long-term prediction variance using the PRE dataset. The long-term prediction variances for the PRE dataset are presented in Figure \ref{GEN-PREC-RECO-VAR}. The proposed method achieves the lowest overall variance throughout the spatiotemporal field, which means that this approach has high robustness for predicting a complex spatiotemporal field. As shown in Figure \ref{GEN-PREC-RECO-VAR}  (a), (d), (e), VAE and RAND-DL achieves a slightly increased reconstruction variance, and Q-DEIM result in a significantly higher variance for all future test snapshots. The future spatiotemporal field reconstruction and prediction of Q-DEIM are significantly less accurate when compared to the proposed method. Hence, although the reconstruction snapshots shown in Figure  \ref{GEN-PREC-RECO} (b) are satisfactory, Q-DEIM is not suitable for long-term spatiotemporal field reconstruction and prediction. Similarly, CS performs poorly in spatiotemporal field reconstruction and prediction, as it produces very high variance in several areas. Besides, as indicated in Figure  \ref{GEN-PREC-RECO} (c), CS cannot produce satisfactory reconstruction. Specifically, CS obtains low MSE and VAR by overfitting the training data, and it is not suitable for reconstructing and predicting the PRE dataset. Last, as indicated in Figure \ref{GEN-PREC-RECO-VAR} (c), RAND results in high long-term prediction variance in almost the entire area. To conclude, the proposed method outperform benchmark algorithms significantly for optimizing the sensor deployment locations in global precipitation reconstruction and prediction.

In summary, the proposed method outperforms the benchmark algorithms significantly in both datasets under all evaluation metrics.

\section{Conclusion} \label{section:futurework_generative}
In this paper, a DL model for spatiotemporal field reconstruction and prediction was developed. Sensor deployment locations over an infinite horizon space were optimized according to the spatiotemporal information and compressed via a measurement matrix. Furthermore, a spatiotemporal reconstructor was established to learn the reconstruction of a spatiotemporal field, based on limited in-situ measurements. In this manner, the sparse sampling locations of the spatiotemporal area were calculated, and the DL model learned the nonlinear mapplings between the sparse representation and the spatiotemporal field. Simulation was conducted using two NOAA datasets. The results showed that the proposed method outperformed the benchmark algorithms in both reconstruction accuracy and long-term prediction roubustness.





\bibliography{bibtex} 
\bibliographystyle{ieeetr}

%





\end{document}